\documentclass[11pt,doublespace,onecolumn]{IEEEtran}

\usepackage[dvips]{epsfig}
\usepackage[dvips]{graphicx}
\usepackage{boxedminipage}
\usepackage{color}
\usepackage{amsfonts}
\usepackage{amssymb}
\usepackage{amsmath}
\usepackage{graphics,amsmath,amsfonts,amssymb,epsfig,latexsym,amsfonts,graphicx,amssymb}
\usepackage{setspace}
\usepackage{cite}

\makeatother  

\newcommand{\trace}{{\mbox{\textrm{Tr}}}}

\newcommand{\st}{{\rm s.t.}}
\newcommand{\by}{\mathbf{y}}

\newcommand{\bx}{\mathbf{x}}

\newcommand{\bJ}{\mathbf{J}}
\newcommand{\bE}{\mathbf{E}}
\newcommand{\bC}{\mathbf{C}}

\newcommand{\bH}{\mathbf{H}}
\newcommand{\bI}{\mathbf{I}}

\newcommand{\bd}{\mathbf{d}}
\newcommand{\bu}{\mathbf{u}}
\newcommand{\bv}{\mathbf{v}}
\newcommand{\bw}{\mathbf{w}}
\newcommand{\bB}{\mathbf{B}}
\newcommand{\bc}{\mathbf{c}}

\newcommand{\cI}{\mathcal{I}}

\newcommand{\cQ}{\mathcal{Q}}

\begin{document}
\title{Joint Base Station Clustering and Beamformer Design for Partial Coordinated Transmission in
Heterogeneous Networks}
\author{ Mingyi Hong, Ruoyu Sun, Hadi Baligh, Zhi-Quan Luo \thanks{Paper received February 1,
2012; revised June 15, 2012. This research is supported in part by
the NSF, Grant No.\ CCF-1216858, and in part by a research gift from
Huawei Technologies Inc.}
\thanks{M.\ Hong, R.\ Sun and
Z.-Q.\ Luo are with the Department of Electrical and Computer
Engineering University of Minnesota, Minneapolis, MN 55455, USA.
Emails: \{mhong, sunxx394, luozq\}@umn.edu.}\thanks{H.\ Baligh is
with Huawei Technologies Canada Co. Ltd., Ottawa, Ontario, Canada.
Email: {hadi.baligh@huawei.com}.}} \maketitle
\begin{abstract}
We consider the interference management problem in a multicell MIMO
heterogeneous network. Within each cell there is a large number of
distributed micro/pico base stations (BSs) that can be potentially
coordinated for joint transmission. To reduce coordination overhead,
we consider user-centric BS clustering so that each user is served
by only a small number of (potentially overlapping) BSs. Thus, given
the channel state information, our objective is to jointly design
the BS clustering and the linear beamformers for all BSs in the
network. In this paper, we formulate this problem from a {sparse
optimization} perspective, and propose an efficient algorithm that
is based on iteratively solving a sequence of group LASSO problems.
A novel feature of the proposed algorithm is that it performs BS
clustering and beamformer design jointly rather than separately as
is done in the existing approaches for partial coordinated
transmission.  Moreover, the cluster size can be controlled by
adjusting a single penalty parameter in the nonsmooth regularized
utility function. The convergence of the proposed algorithm (to a
stationary solution) is guaranteed, and its effectiveness is
demonstrated via extensive simulation.
\end{abstract}

\section{Introduction}
The design of future wireless cellular networks is on the verge of a
major paradigm change. In order to accommodate the explosive demand
for wireless data, the traditional wireless network architecture
comprised of a small number of high power base stations (BSs) has
started to migrate to the so-called heterogeneous network (HetNet)
\cite{3gpp09,damnjanovic11}. In HetNet, each cell is composed of
potentially a large number of densely deployed access nodes such as
macro/micro/pico BSs to provide coverage extension for cell edge and
hotspot users \cite{damnjanovic11}. Unfortunately, close proximity
of many transmitters and receivers introduces substantial
interference, which, if not properly managed, can significantly
affect the system performance.

The interference management problem in multicell downlink networks
has been a topic of intensive research recently. It has been widely
accepted that combining physical layer techniques such as multiple
input multiple output (MIMO) antenna arrays with multi-cell
coordination can effectively mitigate inter-cell and intra-cell
interference \cite{Foschini06, gesbert07, gesbert10}. There are two
main approaches for the coordinated transmission and reception in a
multi-cell MIMO network: joint processing (JP) and coordinated
beamforming (CB) \cite{Foschini06}. In the first approach, the user
data signals are shared among the cooperating BSs. A single virtual
BS can then be formed that transmits to all the users in the system.
Inter-BS interference is canceled by joint precoding and
transmission among all the coordinated BSs. In this case, either the
capacity achieving non-linear dirty-paper coding (DPC) (see, e.g.,
\cite{Caire03,yu07perantenna}), or simpler linear precoding schemes
such as zero-forcing (ZF) (see, e.g., \cite{Spencer04, zhang09,
zhang10JSAC}) can be used for joint transmission. However,
centralized processing is needed for the computation of the
beamformers. Furthermore, this approach can require heavy signaling
overhead on the backhaul network (\!\!\cite{Love08,gesbert10,
Irmer11}) especially when the number of cooperating BSs in the
network becomes large.

When the benefit of full JP among the BSs is outweighed
by the overhead, the BSs can choose CB as an alternative reduced
coordination scheme. In particular, the beamformers
are jointly optimized among the coordinated BSs to suppress
excessive inter-BS interference. In this case, only local channel
state information (CSI) and control information are exchanged among
the coordinated BSs. One popular formulation for CB is to optimize the system
performance measured by a certain system utility function.
Unfortunately, optimally solving the utility maximization problem in  MIMO interfering
network is computationally intractable in general (except for a few exceptions, see
\cite{luo08a, liu11ICC, liu11MISO, Razaviyayn11Asilomar}). As a result, many works are
devoted to finding high quality locally optimal solutions for different
network configurations, e.g., in MIMO/MISO interference channels (IC), \cite{kim11,
shi2009pricingmimo, Ho10,Larsson08,Jorswieck08misogame} and MIMO/MISO interfering
broadcast channels (IBC)
\cite{hong11m_lowerbound, venturino10,shi11WMMSE_TSP}. In particular, reference
\cite{shi11WMMSE_TSP} proposed a weighted Minimum Mean
Square Error (WMMSE) algorithm that is able to compute locally optimal solutions for
a broad class of system utility functions and for general network configurations.

A different approach for limited coordination is to group
the BSs into coordination clusters of small sizes, within which they perform JP.
In this case, each user's data signals are only shared among a small number of its serving BSs,
thus greatly reducing the overall backhaul signaling cost. Many recent works have developed various
BS clustering strategies for such purpose, e.g., \cite{zhang09, Ng10,
Moon11,Kaviani10scheduling, Papadogiannis08clustering, Papadogiannis09, Papadogiannis10, Marsch07},
where clusters are formed either greedily or by an exhaustive search procedure. Once the clusters are formed,
various approaches can be used to design beamforming strategies for each BS. For example, the authors of
\cite{Papadogiannis08clustering, Papadogiannis09, Papadogiannis10}
utilized the ZF strategy for intra-cluster transmission without assuming any inter-cluster cooperation.
References \cite{zhang09, Ng10} considered a hybrid cooperation strategy in which CB is used for
inter-cluster coordination. In this way, inter-cluster interference for cluster edge users is also mitigated.
In principle, clustering strategies should be designed in conjunction
with the beamforming  and BS coordination strategies to
strike the best tradeoff among system throughput performance and
signalling overhead. 

In this work, we consider the joint BS clustering and beamformer
design problem in a downlink multicell HetNet for general partial coordinated transmission.
In our formulation, the BSs that belong to the same cell can dynamically form (possibly overlapping)
coordination clusters of small sizes for JP while the BSs in different cells perform CB.
We formulate this problem from the perspective of sparse
optimization. Specifically, if all the BSs that belong to the same
cell are viewed as a {\it single} virtual BS, then its antennas can
be partitioned into multiple groups (each corresponding to an
individual BS). Moreover, the requirement that each user is served
by a small number of BSs translates directly to the restriction that
its virtual beamformer should have a {\it group sparse} structure,
that is, the nonzero components of the virtual beamformer should
correspond to only a small number of antenna groups. This
interpretation inspires us to formulate a system utility
maximization problem with a mixed $\ell_2/\ell_1$ regularization, as it is
well known that such regularization induces the group sparse
structure \cite{yuan06}. Incorporating such nonsmooth regularization
term into our objective ensures that the optimal beamformers possess
the desired group-sparse structure. In this way, our proposed
approach can be viewed as a {\it single-stage} formulation of the
joint BS grouping and beamforming problem. The main contributions of
this work are listed as follows.
\begin{itemize}
\item We propose to jointly optimize the coordination clusters and linear beamformers in a large
scale HetNet by solving a single-stage nonsmooth utility optimization problem. The system utility
function (without nonsmooth penalization) can have a very general form that includes the popular weighted sum rate
and proportional fair utility functions. This approach is different from
the existing algorithms, which either require a predefined BS clustering and a fixed system utility function,
or some multi-stage heuristic optimization.

\item Since the resulting nonsmooth utility maximization problem is difficult to solve due to its nonconvexity as
well as its nonsmoothness, we transform this problem to an {\it equivalent} regularized weighted MSE minimization
problem. The latter has several desirable features such as separability across the cells and convexity among different
blocks of variables. This equivalence transformation substantially generalizes our previous result
in \cite{shi11WMMSE_TSP}, which only deals with smooth utility functions.
\item We propose an efficient iterative algorithm that computes a stationary solution to the transformed problem.
In each step of the algorithm the computation is closed-form and can be distributed to individual cells.
The algorithm is shown to converge to a stationary solution to the
original nonsmooth utility maximization problem,  and its effectiveness
is demonstrated via extensive simulation experiments.
\end{itemize}

The rest of the paper is organized as follows. In Section
\ref{secSys}, we present the system model and formulate the problem
into a nonsmooth utility maximization problem. We then transform
this problem into an equivalent regularized weighted MSE
minimization problem in Section \ref{secEquivalent}. In Section
\ref{secAlgorithm}, an efficient algorithm is proposed to solve the
transformed problem. In Section \ref{secSimulation}, numerical
examples are provided to validate the proposed algorithm.

{\it Notations}: For a symmetric matrix $\mathbf{X}$,
$\mathbf{X}\succeq 0$ signifies that $\mathbf{X}$ is positive
semi-definite. We use $\trace(\mathbf{X})$, $|\mathbf{X}|$,
$\mathbf{X}^H$ and $\rho(\mathbf{X})$ to denote the trace,
determinant, hermitian and spectral radius of a matrix,
respectively. For a complex scalar $x$, its complex conjugate is
denoted by $\bar{x}$. For a vector $\bx$, we use $\|\bx\|$ to denote
its $\ell_2$ norm. $\mathbf{I}_n$ is used to denote a $n\times n$
identity matrix. We use $[y,\mathbf{x}_{-i}]$ to denote a vector
$\mathbf{x}$ with its $i$th element replaced by $y$. We use
$\mathbb{R}^{N\times M}$ and $\mathbb{C}^{N\times M}$ to denote the
set of real and complex $N\times M$ matrices; We use
$\mathbb{S}^{N}$ and $\mathbb{S}^{N}_{+}$ to denote the set of
$N\times N$ hermitian and hermitian positive semi-definite matrices,
respectively. We use the expression: $0\le a \perp b\ge 0$ to
indicate $a\ge0, b\ge 0, a\times b=0$.

\section{System Model and Problem Formulation}\label{secSys}
Consider a downlink multi-cell HetNet consisting of a set
$\mathcal{K}\triangleq\{1,\cdots,K\}$ of cells. Within each cell $k$
there is a set of $\mathcal{Q}_k=\{1,\cdots,Q_k\}$ distributed base
stations (BS) (for instance, macro/micro/pico BSs) which provide
service to users located in different areas of the cell. Assume that
in each cell $k$, there is low-latency backhaul network connecting
the set of BSs $\mathcal{Q}_k$ to a central controller (usually the
macro BS), and that the central controller makes the resource
allocation decisions for all BSs within the cell. Furthermore, this
central entity has access to the data signals of all the users in
its cell. Let $\mathcal{I}_k\triangleq\{1,\cdots, I_k\}$ denote the
users located in cell $k$. Each of the users $i_k\in\mathcal{I}_k$
is served jointly by a subset of BSs in $\mathcal{Q}_k$. Let
$\mathcal{I}$ denote the set of all the users. For simplicity of
notations, let us assume that each BS has $M$ transmit antennas, and
each user has $N$ receive antennas. Throughout the paper, we use
$i,j$ to indicate the user index, use $k,\ell$ for the cell index,
and use $q,p$ for the BS index. Let
$\mathbf{H}^{q_\ell}_{i_k}\in\mathbb{C}^{N\times M}$ denote the
channel matrix between the $q$th BS in the $\ell$th cell and the
$i$th user in the $k$th cell. Let
$\mathbf{H}^{\ell}_{i_k}\triangleq\left[\mathbf{H}^{1}_{i_k},\cdots,\mathbf{H}^{Q_\ell}_{i_k}\right]\in\mathbb{C}^{N\times
MQ_\ell}$ denote the channel matrix between all the BSs in the
$\ell$th cell to the user $i_k$.

Let $\bv^{q_k}_{i_k}\in\mathbb{C}^{M\times 1}$ denote the transmit
beamformer that BS $q_k$ uses to transmit a single stream of data signal
$s_{i_k}\in\mathbb{C}$ to user $i_k$. Define
$\bv_{i_k}\triangleq\left[(\bv^{1}_{i_k})^H,\cdots,(\bv^{Q_k}_{i_k})^H\right]^H\in\mathbb{C}^{M
Q_k\times 1}$ as 
the collection of all beamformers intended for user $i_k$. Let
$\bv\triangleq\left[\bv^H_{1},\cdots,\bv^H_{I_k}\right]^H$. Assume
that there is a power budget constraint for each BS $q_k$,
i.e.,{\small
\begin{align}
\sum_{i_k\in\mathcal{I}_k}(\bv^{q_k}_{i_k})^H\bv^{q_k}_{i_k}\le
P_{q_k},\ \forall~q_k\in\mathcal{Q}_k,~\forall~k\in\mathcal{K}.
\end{align}}\hspace{-0.2cm}
Let $\mathbf{x}^{q_k} \in \mathbb{C}^{M\times 1}$ denote the
transmitted signal of BS $q_k$, and let
$\mathbf{x}^k\triangleq\left[(\bx^{1})^H,\cdots,(\bx^{Q_k})^H\right]^H\in\mathbb{C}^{M
Q_k\times 1}$ denote the collection of transmitted signals of all
the BSs in cell $k$, i.e.
\begin{align}
{\bx}^{q_k}=\sum_{i_k\in\mathcal{I}_k}\bv^{q_k}_{i_k}s_{i_k},\quad\quad
{\bx}^{k}=\sum_{i_k\in\mathcal{I}_k}\bv_{i_k}s_{i_k}\nonumber.
\end{align}
The received signal $\by_{i_k}\in\mathbb{C}^{N\times 1}$ of user
$i_k$ is{\small
\begin{align}
\by_{i_k}&=\sum_{\ell\in\mathcal{K}}\bH^{\ell}_{i_k}\bx^{\ell}+\mathbf{z}_{i_k}\nonumber\\
&= \bH^{k}_{i_k}\bv_{i_k}s_{i_k}+\underbrace{\sum_{j_k\ne
i_k}\bH^{k}_{i_k}\bv_{j_k}s_{j_k}}_{\textrm{intra-cell
interference}}+\underbrace{\sum_{\ell\ne
k}\sum_{j_\ell\in\mathcal{I}_\ell}\bH^{\ell}_{i_k}\bv_{j_\ell}s_{j_\ell}}_{\textrm{inter-cell
interference}}+\mathbf{z}_{i_k}
\end{align}}\hspace{-0.2cm}
where $\mathbf{z}_{i_k}\in\mathbb{C}^{N\times 1}$ is the additive white Gaussian
noise with distribution
$\mathcal{CN}(0,\sigma^2_{i_k}\mathbf{I}_N)$.

Let $\bu_{i_k}\in\mathbb{C}^{N\times 1}$ denote the receive
beamformer used by user $i_k$ to decode the intended signal. Then
the estimated signal for user $i_k$ is:
$\widehat{s}_{i_k}=\bu^H_{i_k}\by_{i_k}$. The mean square error
(MSE) for user $i_k$ can be written as{\small
\begin{align}
e_{i_k}&\triangleq
\bE[(s_{i_k}-\widehat{s}_{i_k})(\overline{s_{i_k}}-\overline{\widehat{s}_{i_k}})]\nonumber\\
&=(1-\bu^H_{i_k}\bH^{k}_{i_k}\bv_{i_k})(1-\overline{\bu^H_{i_k}\bH^{k}_{i_k}\bv_{i_k}})\nonumber\\
&\quad +\sum_{(\ell,j)\ne
(k,i)}\bu^H_{i_k}\bH^\ell_{i_k}\bv_{j_\ell}\bv^H_{j_\ell}(\bH^\ell_{i_k})^H\bu_{i_k}+\sigma^2_{i_k}\bu^H_{i_k}\bu_{i_k}.\label{eqMSE}
\end{align}}\hspace{-0.2cm}
The MMSE receiver minimizes user $i_k$'s MSE, and can be expressed as{\small
\begin{align}
\bu^{\rm{mmse}}_{i_k}&=\left(\sum_{(\ell,j)}\bH^{\ell}_{i_k}\bv_{j_\ell}(\bv_{j_\ell})^H(\bH^{\ell}_{i_k})^H
+\sigma^2_{i_k}\bI\right)^{-1}\bH^{k}_{i_k}\bv_{i_k}\nonumber\\
&\triangleq
\bC^{-1}_{i_k}\bH^{k}_{i_k}\bv_{i_k}\label{eqDefCCapital}
\end{align}}\hspace{-0.2cm}
where $\bC_{i_k}$ denotes user $i_k$'s received signal covariance
matrix. The minimum MSE for user $i_k$ when the MSE receiver is used
can be expressed as{\small
\begin{align}
e^{\rm
mmse}_{i_k}=1-(\bv_{i_k})^H(\bH^k_{i_k})^H\mathbf{C}^{-1}_{i_k}
\bH^k_{i_k} \bv_{i_k}.\label{eqMMSE}
\end{align}}\hspace{-0.15cm}
Clearly, we have $1-e^{\rm mmse}_{i_k}\ge 0$. Let us assume that
Gaussian signaling is used and the interference is treated as noise.
If we assume that all the BSs in cell $k$ form a single virtual BS,
then $\bv_{i_k}$ can be viewed as the virtual beamformer for user
$i_k$. The achievable rate for user $i_k$ is given by
\cite{cover05}{\small
\begin{align}
R_{i_k}&=\log\bigg|\mathbf{I}_{N}+\mathbf{H}^k_{i_k}\mathbf{v}_{i_k}\bv^H_{i_k}({\bH}^k_{i_k})^H\nonumber\\
&\quad \times\big(\sum_{(\ell,j)\ne (k,i)}
\mathbf{H}^\ell_{i_k}\mathbf{v}_{j_\ell}\bv^H_{j_\ell}({\bH}^{\ell}_{i_k})^H+\sigma^2_{i_k}\mathbf{I}_N\big)^{-1}\bigg|\label{eqRate}.
\end{align}}\hspace{-0.2cm}
The above expression suggests that each user can always use a MMSE receiver
\eqref{eqDefCCapital} since it preserves achievable data rate
when the interference is treated simply as noise. We will
occasionally use the notations $R_{i_k}(\bv)$, $\bC_{i_k}(\bv)$ to
make their dependencies on $\bv$ explicit.


Notice that the rate \eqref{eqRate} can only be achieved when all
the BSs in cell $k$ perform a full JP. Unfortunately, this requires
the data signal for user $i_k$ to be known at all BSs in ${\cal
Q}_k$, causing significant signaling overhead, especially when the
number of users and BSs becomes large. To reduce overhead, partial
cooperative transmission is preferred whereby each user is served by
not all, but a subset, of BSs in the cell. Mathematically, we are
interested in jointly performing the following two tasks: {\it i)}
for each user $i_k$, identify a small subset of serving BSs
$\mathcal{S}_{i_k}\subseteq\mathcal{Q}_{k}$ such that for each
$q_k\notin\mathcal{S}_{i_k}$, $\bv^{q_k}_{i_k}=\mathbf{0}$; {\it
ii)} optimize the transmit beamformers
$\{\bv_{i_k}^{q_k}\}_{q_k\in\mathcal{S}_{i_k}, i_k\in\mathcal{I}_k}$
to achieve high system throughput and/or fairness level. With the
partial JP, user $i_k$'s data signal needs to be shared only among
BSs in ${\cal S}_{i_k}$, rather than among all BSs in ${\cal Q}_k$.

The requirement that $|\mathcal{S}_{i_k}|$ is small translates to
the restriction that $\bv_{i_k}$ should contain only a few nonzero
block components (i.e., most of the beamformers
$\{\bv^{q_k}_{i_k}\}_{q_k\in\mathcal{Q}_k}$ should be set to zero).
This structure of the beamformer $\bv_{i_k}$ is referred to as the
{\it group sparse} structure \cite{yuan06}. Recovering group sparse
solutions for optimization problem has recently found its
application in many fields such as machine learning \cite{Bach07},
microarray data analysis \cite{Ma07supervisedgroup}, signal
processing \cite{Wright09, Eldar11} and communications
\cite{Bazerque11}. One popular approach to enforce the sparsity of
the solution to an optimization problem is to penalize the objective
function with a group-sparse encouraging penalty such as the mixed
$\ell_2/\ell_1$ norm \cite{yuan06}. In our case, such norm can be
expressed as: $\sum_{q_k\in\mathcal{Q}_k}\|\bv^{q_k}_{i_k}\|$, which
is the $\ell_1$ norm of the vector consists of $\ell_2$ norms $\{
\|\bv_{i_k}^{{q}_k}\|\}_{q_k\in\mathcal{Q}_k}$. The resulting
penalized problem is usually referred to as the {group
least-absolute shrinkage and selection operator} (LASSO) problem.

With the goal of inducing group-sparse structure of the beamformers
$\{\bv_{i_k}\}_{i_k\in\mathcal{I}}$ as well as optimizing
system-level performance, we propose to design the linear transmit
beamformers by solving the following problem {\small
\begin{align}
{\rm
(P1)}&\quad\max_{\{\bv^{q_k}_{i_k}\}}\sum_{k\in\mathcal{K}}\sum_{i_k\in\mathcal{I}_k}\left(u_{i_k}(R_{i_k})
-\lambda_k\sum_{q_k\in\mathcal{Q}_k}\|\bv^{q_k}_{i_k}\|\right)\label{problemGeneral}\\
\st&\quad
\sum_{i_k\in\mathcal{I}_k}(\bv^{q_k}_{i_k})^H\bv^{q_k}_{i_k}\le
P_{q_k},\
\forall~q_k\in\mathcal{Q}_k,~\forall~k\in\mathcal{K}\nonumber
\end{align}}\hspace{-0.2cm}
where $u_{i_k}(\cdot)$ denotes user $i_k$'s utility function, and
$\{\lambda_k\ge 0\}_{k\in\mathcal{K}}$ is the set of parameters that
control the level of sparsity within each cell. Penalizing the
objective with the nonsmooth mixed $\ell_2/\ell_1$ norm induces the
group sparsity of $\bv_k$. To see this, note that the group sparsity
of $\bv_k$ can be characterized by the sparsity of the vector $\{
\|\bv_{i_k}^{{q}_k}\|\}_{q_k\in\mathcal{Q}_k}$, and the $\ell_1$
norm of this vector, which is what we use in \eqref{problemGeneral},
is a good approximation of the $\ell_0$ norm of it (defined as the
number of nonzero entries of the vector); see the literature on
compressive sensing \cite{Donoho06}. The reference  \cite{yuan06}
contains more discussion on using the mixed $\ell_2/\ell_1$ norm to
recover the group sparsity.

Unfortunately, solving  ${\rm (P1)}$ directly is challenging. One
reason is that when $\lambda_k = 0, \forall k$, (P1) becomes a sum
utility maximization problem for an interfering broadcast channel,
which is proven to be NP-hard for many common utility functions (see
\cite{luo08a, liu11ICC, liu11MISO, Razaviyayn11Asilomar}). Another
reason is that most existing algorithms for solving the nonsmooth
group LASSO problem such as \cite{yuan06, qin11m,deng11m,Wright09}
only work for the case that the smooth part of the objective is {\it
convex and quadratic}. We will provide in the next section an
equivalent reformulation of this problem in which these difficulties
are circumvented. The reformulated problem can be solved (to a
stationary solution) via solving a series of convex problems.

\section{Equivalent Formulation}\label{secEquivalent}

In this section, we develop a general equivalence relationship
between the utility maximization problem ${\rm (P1)}$ and a
regularized weighted MSE minimization problem. This result is a
generalization of a recent equivalence relationship developed in
\cite{shi11WMMSE_TSP} to the nonsmooth setting. The proofs of the
results in this section can be found in Appendix
\ref{appEquivalence}.

\subsection{Single User Per Cell with Sum Rate Utility}\label{subEquivalentSingle}
For ease of presentation, we first consider a simpler case in which
there is a single user in each cell. This scenario is of interest
when different mobiles in each cell are scheduled to orthogonal
time/frequency resources, and we consider one of such resources. We
also focus on using the sum rate utility function. Generalizations
to multiple users per-cell case with more general utility functions
will be given in the next subsection.

Now that there is a single user in each cell, we denote the user in
$k$th cell as user $k$. We use $\bv^{q_k}_k$ and $\bH^{q_\ell}_k$ to
denote the BS $q_k$'s beamformer for user $k$, and the channel from
BS $q_\ell$ to user $k$, respectively. Define $R_k$, $e_k$, $\bv_k$,
$\bH^{\ell}_k$ and
$\bu_k$ similarly. 
Using the sum rate as the system utility function, the sparse
beamforming problem for this network configuration is given as{\small
\begin{align}
\max_{\{\bv^{q_k}_{k}\}}&\sum_{k\in\mathcal{K}}\bigg(R_{k}
-\lambda_k\sum_{q_k\in\mathcal{Q}_k}\|\bv^{q_k}_{k}\|\bigg)\label{problemSingle}\\
\st\quad &(\bv^{q_k}_{k})^H\bv^{q_k}_{k}\le P_{q_k},\
\forall~q_k\in\mathcal{Q}_k,~\forall~k\in\mathcal{K}\nonumber.
\end{align}}\hspace{-0.2cm}
Let us introduce a set of new weight variables
$\{w_k\}_{k\in\mathcal{K}}$. Consider the following {\it regularized
weighted MSE minimization problem} {\small
\begin{align}
\min_{\{\bv^{q_k}_{k}\}, \{\bu_k\},
\{w_k\}}&\sum_{k\in\mathcal{K}}\bigg(w_k e_k-\log(w_k)+\lambda_k\sum_{q_k\in\mathcal{Q}_k}\|\bv^{q_k}_k\|\bigg)\label{problemWMMSESingle}\\
\st\quad& (\bv^{q_k}_{k})^H\bv^{q_k}_{k}\le P_{q_k},\
\forall~q_k\in\mathcal{Q}_k,~\forall~k\in\mathcal{K}\nonumber
\end{align}}\hspace{-0.2cm}
One immediate observation is that fixing $\bv,\bu$ and solving for
$\bw$ admits  a closed form solution:
$w_k=\frac{1}{{e}_k},~\forall~k$. Such property will be used in the
following to derive the equivalence relationship between problems
\eqref{problemSingle} and \eqref{problemWMMSESingle}. We refer the
readers to \cite[Section II.A]{shi11WMMSE_TSP} for a simple example
that motivates this equivalence in the case $\lambda_k=0$.

To formally derive the equivalence relationship, the following
definitions (see \cite{tseng01}) of stationary
points of a nonsmooth function are needed. Note that stationarity
is a necessary condition for both global and local
optimality. Let $\mathbf{x}=[\mathbf{x}_1,\cdots,\mathbf{x}_K]$ be a
vector of variables, in which $\mathbf{x}_k\in\mathbb{C}^{N_k}$. Let
$f(\cdot): \mathbb{C}^{\sum_{k}N_k\times 1}\to \mathbb{R}$ be a real
valued (possibly nonsmooth) continuous function.

\newtheorem{D1}{\bf Definition}
\begin{D1}\label{defStationary}
{\it $\bx^*$ is a stationary point of the problem $\min
f(\mathbf{x})$ if $\bx^* \in{\rm{dom}}(f)$ and $f'(\bx^*;\bd)\ge 0, \
\forall~\bd$, where $f'(\bx^*;\bd)$ is the directional derivative of
$f(\cdot)$ at $\bx^*$ in the direction $\bd${\small
\begin{align}
f'(\bx^*;\bd)=\lim\inf_{\lambda \downarrow
0}[f(\bx^*+\lambda\bd)-f(\bx^*)]/\lambda.\nonumber
\end{align}}\hspace{-0.2cm}}
\end{D1}
\vspace{-0.6cm}
\newtheorem{D2}{Definition}
\begin{D1}\label{defCoordinateMin}
{\it $\bx^*$ is a coordinatewise stationary point of $\min
f(\mathbf{x})$ if $\bx^* \in{\rm{dom}}(f)$ and{\small
\begin{align}
&f(\bx^*+[0,\cdots,0,\bd_k,0,\cdots,0])\ge f(\bx^*), \
\forall~\bd_k\in\mathbb{C}^{N_k},\
\nonumber\\
&\quad \forall~\bx^*+[0,\cdots,0,\bd_k,0,\cdots,0]\in{\rm
{dom}}(f),~\forall~k=1,\cdots,K\nonumber
\end{align}}\hspace{-0.2cm}
where $[0,\cdots,0,\bd_k,0,\cdots,0]$ denotes a vector with all zero
components except for its $k$th block.}
\end{D1}

\newtheorem{D3}{Definition}
\begin{D1}\label{defRegular}
{\it The function $f(\cdot)$ is regular at $\bx^* \in{\rm{dom}}(f)$ if
$f'(\bx^*;(0,\cdots,0,\bd_k,0,\cdots,0))\ge 0,~
\forall~\bd_k\in\mathbb{C}^{N_k},~\forall~k=1,\cdots,K$ implies
$f'(\bx^*,\bd)\ge 0, \ \forall~\bd=[\bd_1,\cdots,\bd_K]$.}
\end{D1}

We establish the equivalence between problems \eqref{problemSingle} and
\eqref{problemWMMSESingle} in the following proposition.
\newtheorem{P1}{\bf Proposition}
\begin{P1}\label{propEquivalence}
{\it If $(\bv^*,\bu^*,\bw^*)$ is a stationary solution to problem
\eqref{problemWMMSESingle}, then $\bv^*$ must be a stationary
solution to problem \eqref{problemSingle}.  Conversely, if $\bv^*$
is a stationary solution to problem \eqref{problemSingle}, then the
tuple $(\bv^*,\bu^*,\bw^*)$  must be a stationary solution to
problem \eqref{problemWMMSESingle}, where{\small
\begin{align}
\bu^*_k&=\mathbf{C}^{-1}_k(\bv^*)\bH^{k}_k\bv^*_k,\
w^*_k\nonumber\\
&=\left(1-(\bv^*_{k})^H(\bH^k_k)^H\mathbf{C}^{-1}_k(\bv^{*})
\bH^k_k \bv^*_{k}\right)^{-1},\ \forall \ k\in\mathcal{K}\nonumber\\
{\rm with}\quad  &\bC_k(\bv^*)\triangleq
\sum_{\ell\in\mathcal{I}}\bH^{\ell}_{k}\bv^*_{\ell}(\bv^*_{\ell})^H(\bH^{\ell}_{k})^H
+\sigma^2_{k}\bI\nonumber.
\end{align}}\hspace{-0.2cm}
Moreover, the global optimal solutions $\bv^*$ for these two
problems are identical. }
\end{P1}

Notice that  $\bu^*_k$ and $w^*_k$ introduced in
Proposition \ref{propEquivalence} are the MMSE receiver and the
inverse MMSE corresponding to the transmit beamformer $\bv^*$
(cf. \eqref{eqDefCCapital} and \eqref{eqMMSE}) respectively.

\vspace{-0.3cm}
\subsection{Multiple Users Per Cell with More General Utility}\label{subEquivalentGeneral}
In this section, we generalize the equivalence relationship
presented in the previous section to the case with more general
utility function and multiple users per cell.

Consider the utility function $u_{i_k}(R_{i_k})$ that satisfies the
following two conditions:
\begin{description}
\item [\bf C1)] $u_{i_k}(x)$ is concave and strictly increasing
 in $x$;
\item [\bf C2)] $u_{i_k}(-\log(x))$ is strictly convex in $x$ for all $x$ satisfying $1\ge x\ge 0$.
\end{description}

Note that this family of utility functions includes several well
known utilities such as weighted sum rate and geometric mean of one
plus rates \cite{shi11WMMSE_TSP}. 
Let $\{w_{i_k}\}_{i_k\in\mathcal{I}}$ be a set of real-valued
weights. Let $\gamma_{i_k}(\cdot): \mathbb{R}\to\mathbb{R}$ denote
the {\it inverse function} of the derivative $\frac{ d [-u_{i_k}(-
\log(e_{i_k}))]}{d e_{i_k}}$. Consider the following  {\it
regularized weighted MSE minimization} problem {\small
\begin{align}
{\rm (P2)}&\min_{\{\bv^{q_k}_{i_k}\},
\{\bu_{i_k}\},\{w_{i_k}\}}\sum_{k\in\mathcal{K}}
\bigg(\sum_{i_k\in\mathcal{I}_k}\Big(w_{i_k}
e_{i_k}-u_{i_k}(-\log(\gamma_{i_k}(w_{i_k})))\nonumber\\
&\quad\quad-w_{i_k}\gamma_{i_k}(w_{i_k})
+\lambda_k\sum_{q_k\in\mathcal{Q}_k}\|\bv^{q_k}_{i_k}\|\Big)\bigg)\nonumber\\
&\st\quad
\sum_{i_k\in\mathcal{I}_k}(\bv^{q_k}_{i_k})^H\bv^{q_k}_{i_k}\le
P_{q_k},\
\forall~q_k\in\mathcal{Q}_k,~\forall~k\in\mathcal{K}\nonumber\\
&\quad \quad e_{i_k} \ \textrm{defined in \eqref{eqMSE}}.\nonumber
\end{align}}
Similar to Proposition \ref{propEquivalence}, we can establish the
following equivalence relationship.
\newtheorem{P2}{Proposition}
\begin{P1}\label{propEquivalenceGeneral}
{\it Suppose for each $i_k\in\mathcal{I}_k$, the utility function $u_{i_k}(\cdot)$ satisfies the
conditions C1)--C2). If $(\bv^*,\bu^*,\bw^*)$ is a stationary solution to problem
${\rm (P2)}$, then $\bv^*$ must be a stationary
solution to problem ${\rm (P1)}$.  Conversely, if $\bv^*$
is a stationary solution to problem ${\rm (P1)}$, then the
tuple $(\bv^*,\bu^*,\bw^*)$  must be a stationary solution to
problem ${\rm (P2)}$, where{\small
\begin{align}
\bu^*_{i_k}&=\mathbf{C}^{-1}_{i_k}(\bv^*)\bH^{k}_{i_k}\bv^*_{i_k},
\nonumber\\
w^*_{i_k}&=\alpha^*_{i_k}\left(1-(\bv^*_{i_k})^H(\bH^k_{i_k})^H\mathbf{C}^{-1}_{i_k}(\bv^*)
\bH^k_{i_k} \bv^*_{i_k}\right)^{-1}\nonumber\\
&{\rm
with~}\alpha^*_{i_k}\triangleq\frac{du_{i_k}(R_{i_k})}{dR_{i_k}}\bigg|_{R_{i_k}=R_{i_k}(\bv^*)}\ge
0\nonumber.
\end{align}}
Moreover, the global optimal solutions $\bv^*$ for these two
problems are identical. }
\end{P1}

We remark that $\bu^*_{i_k}$ is again the MSE receiver corresponding
to $\bv^*$; $w^*_{i_k}$ takes a similar form to that in the
statement of Proposition \ref{propEquivalence}, except for the
inclusion of a positive weight $\alpha^*_{i_k}$. The positivity of
$\alpha^*_{i_k}$ comes form the assumption C1).

\vspace{-0.3cm}
\section{Joint Clustering and Beamformer Design}\label{secAlgorithm}
In this section, we will develop an efficient iterative algorithm
for the general nonsmooth utility maximization problem ${\rm (P1)}$.
Due to the equivalence of this problem and the regularized weighted
sum-MSE minimization problem ${\rm (P2)}$, we can focus on solving
the latter. We will employ the block coordinate descent (BCD) method
\cite{tseng01} for such purpose.

\vspace{-0.3cm}
\subsection{The Algorithm}\label{subAlgorithm}
It is straightforward to verify that the objective of problem
${\rm (P2)}$ is convex w.r.t. each variable
$\bv,\bu,\bw$. When $\bv,\bw$ are fixed, the optimal $\bu^*$ is the
MMSE receiver $
\bu^*_{i_k}=\bC^{-1}_{i_k}\bH^k_{i_k}\bv_{i_k},\ \forall \ i_k\in\mathcal{I}$.
When $\bv,\bu$ are fixed, the optimal $\bw^*$ takes
the following form{\small
\begin{align}
w^*_{i_k}&=\frac{du_{i_k}(R_{i_k})}{dR_{i_k}}\bigg|_{R_{i_k}=R_{i_k}(\bv)}
\times\frac{1}{e_{i_k}}> 0, \ \forall\ i_k\in\mathcal{I}\nonumber.
\end{align}}\hspace{-0.2cm}
The positivity of
$w^*_{i_k}$ comes from the fact that $e_{i_k}> 0$ and the utility
function $u_{i_k}(\cdot)$ is strictly increasing w.r.t. user $i_k$'s
rate.

The main part of the algorithm is to find the optimal transmit
beamformers $\bv$ when $\bu,\bw$ are fixed. 
Observe that when fixing $\bu,\bw$, problem ${\rm (P2)}$ can be decomposed into
$K$ independent convex problems (one for each cell)
{\small
\begin{align}
{\rm (P3)}\min_{\{\bv_{i_k}\}_{i_k\in\mathcal{I}_k}}&\sum_{i_k\in\mathcal{I}_k}\bigg(\bv^H_{i_k}
\Big(\sum_{j_l\in\mathcal{I}}w_{j_l}
(\bH^k_{j_l})^H\bu^H_{j_l}\bu_{j_l}\bH^k_{j_l}\Big)\bv_{i_k}\nonumber\\
&\hspace{-1cm}-w_{i_k}\bv^H_{i_k}(\bH^k_{i_k})^H\bu_{i_k}-w_{i_k}\bu^H_{i_k}\bH^k_{i_k}\bv_{i_k}
+\lambda_k\sum_{q_k\in\mathcal{Q}_k}\|\bv^{q_k}_{i_k}\|\bigg)\nonumber\\
\st&\quad
\sum_{i_k\in\mathcal{I}_k}(\bv^{q_k}_{i_k})^H\bv^{q_k}_{i_k}\le
P_{q_k},\ \forall~q_k\in\mathcal{Q}_k\nonumber.
\end{align}}\hspace{-0.2cm}
Let us focus on solving one of such problems. Note that the constraint set of
this problem is
{separable} among the beamformers of different BSs. This
suggests that we can obtain its optimal solution again by a BCD
method, with $\{\bv^{q_k}_{i_k}\}_{i_k\in\mathcal{I}_k}$ as one block of variables.
In particular, we will solve {\rm (P3)}
by sequentially solving the following problem for each block $\{\bv^{q_k}_{i_k}\}_{i_k\in\mathcal{I}_k}$
{\small
\begin{align}
{\rm (P4)}\min_{\{\bv^{q_k}_{i_k}\}_{i_k\in\mathcal{I}_k}}&\sum_{i_k\in\mathcal{I}_k}\bigg(\bv^H_{i_k}
\Big(\sum_{j_l\in\mathcal{I}}w_{j_l}
(\bH^k_{j_l})^H\bu^H_{j_l}\bu_{j_l}\bH^k_{j_l}\Big)\bv_{i_k}\nonumber\\
&\hspace{-1cm}-w_{i_k}\bv^H_{i_k}(\bH^k_{i_k})^H\bu_{i_k}-w_{i_k}\bu^H_{i_k}\bH^k_{i_k}\bv_{i_k}
+\lambda_k\|\bv^{q_k}_{i_k}\|\bigg)\nonumber\\
\st&\quad
\sum_{i_k\in\mathcal{I}_k}(\bv^{q_k}_{i_k})^H\bv^{q_k}_{i_k}\le
P_{q_k}\nonumber.
\end{align}}\hspace{-0.2cm}
This problem is a quadratically {constrained} group-LASSO problem.
The presence of the additional sum power constraint prevents the
direct application of the algorithms (e.g., \cite{yuan06,
qin11m,deng11m,Wright09}) for conventional unconstrained group-LASSO
problem. In the following we will derive a customized algorithm for
solving this problem.

Define the following two sets of variables{\small
\begin{align}
\bJ_k&\triangleq\sum_{j_l\in\mathcal{I}}w_{j_l}(\bH^{k}_{j_l})^H\bu_{j_l}\bu^H_{j_l}\bH^{k}_{j_l}\in\mathbb{S}_{+}^{MQ_k}\label{eqJGeneral}\\
\bd_{i_k}&\triangleq w_{i_k}(\bH^k_{i_k})^H\bu_{i_k}\in\mathbb{C}^{MQ_k\times
1},\ \forall~i_k\in\mathcal{I}_k.\label{eqDGeneral}
\end{align}}\hspace{-0.2cm}
Partition $\bJ_k$ and $\bd_{i_k}$ into the following form{\small
\begin{align}
\bJ_k&=\left[ \begin{array}{lll}
\bJ_k[1,1],&\cdots,&\bJ_k[1,Q_k]\\
\vdots&\ddots&\vdots\\
\bJ_k[Q_k,1]&\cdots&\bJ_k[Q_k,Q_k]
\end{array}\right],\nonumber\\
\bd_{i_k}&=\left[\bd_{i_k}^H[1],\cdots,\bd_{i_k}^H[Q_k]\right]^H\label{eqPartitionDSingle}
\end{align}}\hspace{-0.2cm}
where $\bJ_k[q,p]\in\mathbb{C}^{M\times M},\
\forall~(q,p)\in\mathcal{Q}_k\times\mathcal{Q}_k$, and
$\bd_{i_k}[q]\in\mathbb{C}^{M\times 1}, \forall~q\in\mathcal{Q}_k$.
Utilizing these definitions, the
objective of problem {\rm (P4)} reduces to{\small
\begin{align}
&\sum_{i_k\in\mathcal{I}_k}(\bv^H_{i_k}\bJ_k\bv_{i_k}-\bv^H_{i_k}\bd_{i_k}-\bd^H_{i_k}\bv_{i_k})+
\lambda_{k}\sum_{i_k\in\mathcal{I}_k}\|\bv^{q_k}_{i_k}\|\nonumber\\
&\triangleq\sum_{i_k\in\mathcal{I}_k}f_{i_k}(\bv_{i_k})+
\lambda_{k}\sum_{i_k\in\mathcal{I}_k}\|\bv^{q_k}_{i_k}\|\nonumber.
\end{align}}\hspace{-0.2cm}
The gradient of the smooth function $f_{i_k}(\bv_{i_k})$ w.r.t. $\bv^{q_k}_{i_k}$ can be
expressed as{\small
\begin{align}
\triangledown_{\bv^{q_k}_{i_k}}{f_{i_k}(\bv_{i_k})}&
=2\bigg(\bJ_k[q,q]\bv^{q_k}_{i_k}+ \sum_{p\ne
q}\bJ_k[q,p]\bv_{i_k}^{p_k}-\bd_{i_k}[q]\bigg)\label{eqGradientVGeneral}\\
&\triangleq 2\bigg(\bJ_k[q,q]\bv^{q_k}_{i_k}-\bc_{i_k}\bigg)
\end{align}}\hspace{-0.2cm}
where we have defined{\small
\begin{align}
\bc_{i_k}\triangleq\bd_{i_k}[q]-\sum_{p\ne
q}\bJ_k[q,p]\bv^{p_k}_{i_k}.\label{eqDefC}
\end{align}}\hspace{-0.2cm}
Note that the gradient \eqref{eqGradientVGeneral} given above is
coupled with other block variables
$\{\bv_{i_k}^{p_k}\}_{i_k\in\mathcal{I}_k}$, $p_k\ne q_k$ through
the term $\sum_{p\ne q}\bJ_k[q,p]\bv^{p_k}_{i_k}$.

The first order optimality condition for the convex problem {\rm
(P4)} is {\small
\begin{align}
&-2\bigg(\bv^{q_k}_{i_k}\mu^{q_k}+ (\bJ_k[q,q]\bv^{q_k}_{i_k}-
\bc_{i_k})\bigg)\in\lambda_k\partial(\|\bv^{q_k}_{i_k}\|),
\ \forall\ i_k\in\mathcal{I}_k,\label{eqLagrangianGeneral}\\
&0\le\mu^{q_k}\perp(P_{q_k}-\sum_{i_k\in\mathcal{I}_k}(\bv^{q_k}_{i_k})^H\bv^{q_k}_{i_k})\ge
0,\label{eqComplimentarityGeneral}
\end{align}}\hspace{-0.2cm}
where $\mu^{q_k}$ is the Lagrangian multiplier for BS $q_k$'s
power budget constraint;  $\partial(\|\bv^{q_k}_{i_k}\|)$ represents
the subdifferential of the nonsmooth function $\|\cdot\|$ at the
point $\bv^{q_k}_{i_k}$. The latter can be expressed as
follows (see \cite{yuan06, luo11note}){\small
\begin{align}
\partial(\|\bv^{q_k}_{i_k}\|)=\left\{ \begin{array}{ll}
\frac{\bv^{q_k}_{i_k}}{\|\bv^{q_k}_{i_k}\|},&\bv^{q_k}_{i_k}\ne \mathbf{0},\\
\left\{\mathbf{x}~|~\|\mathbf{x}\|\le
1\right\},&\bv^{q_k}_{i_k}=\mathbf{0}.
\end{array}\right.\label{eqSubL2}
\end{align}}\hspace{-0.2cm}
Finding the global optimal solution of problem
{\rm (P4)} amounts to finding the optimal
primal dual pair $\{(\bv^{q_k}_{i_k})^*\}_{i_k\in\mathcal{I}_k},(\mu^{q_k})^*$ that satisfy the
conditions \eqref{eqLagrangianGeneral}--\eqref{eqComplimentarityGeneral}. 
In the following, we will first develop a procedure to find
$\{\bv^{q_k}_{i_k}\}_{i_k\in\mathcal{I}_k}$ that satisfy the conditions
\eqref{eqLagrangianGeneral} for a given $\mu^{q_k}\ge 0$. Then we
will use a bisection method to search for the optimal multiplier.

{\bf Step 1)} Utilizing the expression for the subdifferential in
\eqref{eqSubL2}, the optimality condition
\eqref{eqLagrangianGeneral} can be rewritten as \cite{yuan06,
luo11note}{\small
\begin{align}
\bv^{q_k}_{i_k}&={\bf 0},~{\rm{if}}~ \big\|\bc_{i_k}\big\|\le \frac{\lambda_k}{2},\label{eqVForceToZero}\\
\bv^{q_k}_{i_k}&=\bigg(\bJ_k[q,q]+\left(\frac{\lambda_k\delta_{i_k}^{q_k}}{2}+\mu^{q_k}\right)\bI_M\bigg)^{-1}
\bc_{i_k},~\textrm{otherwise}\label{eqVGreaterThanZero}
\end{align}}\hspace{-0.2cm}
with $\delta_{i_k}^{q_k}>0$ defined as
$\frac{1}{\delta_{i_k}^{q_k}}\triangleq\|\bv^{q_k}_{i_k}\|$. Note
that \eqref{eqVForceToZero} is the key to achieve sparsity, as
whenever $\|\bc_{i_k}\|$ is less than the threshold $\lambda_k/2$,
$\bv^{q_k}_{i_k}$ will be forced to $\bf{0}$. The correctness of \eqref{eqVGreaterThanZero}
can be checked by plugging the second part of \eqref{eqSubL2} into
\eqref{eqLagrangianGeneral}.

By definition, the optimal $\delta_{i_k}^{q_k}$ must satisfy {\small
\begin{align}
h_{i_k}(\delta_{i_k}^{q_k},\mu^{q_k})&\triangleq\delta_{i_k}^{q_k}\bigg\|\bigg(\bJ_k[q,q]+(\frac{\lambda_k\delta^{q_k}_{i_k}}{2}
+\mu^{q_k})\bI_M\bigg)^{-1}\bc_{i_k}\bigg\|=1\label{eqDeltaEquality}.
\end{align}}\hspace{-0.2cm}
Define the set of {\it active users} for BS $q_k$ as
$\mathcal{A}^{q_k}\triangleq\big\{i_k|i_k\in\mathcal{I}_k,
\big\|\bc_{i_k}\big\|> \frac{\lambda_k}{2}\big\}$, and define its
cardinality as $|\mathcal{A}^{q_k}|=A^{q_k}$. For any given
$\mu^{q_k}\ge 0$, let us denote a beamformer $\bv^{q_k}_{i_k}$ that
satisfies \eqref{eqVForceToZero}-\eqref{eqVGreaterThanZero} as
$\bv^{q_k}_{i_k}(\mu^{q_k})$, and the corresponding
$\delta_{i_k}^{q_k}$ that satisfies \eqref{eqDeltaEquality} as
$\delta_{i_k}^{q_k}(\mu^{q_k})$. Clearly for a user $i_k$ that
satisfies the condition \eqref{eqVForceToZero} (i.e.,
$i_k\in\mathcal{I}_k\setminus\mathcal{A}^{q_k}$ ),
$\bv^{q_k}_{i_k}(\mu^{q_k})$ does not depend on $\mu^{q_k}$ and can
be directly computed.  Let us then focus on the {\it active user}
$i_k\in\mathcal{A}^{q_k}$. For any $i_k\in\mathcal{A}^{q_k}$,
finding a $\bv_{i_k}^{q_k}(\mu^{q_k})$ amounts to obtaining the
corresponding $\delta_{i_k}^{q_k}(\mu^{q_k})$ that satisfies
\eqref{eqDeltaEquality}. Due to a certain monotonicity property of the
function $h_{i_k}(\delta_{i_k}^{q_k},\mu^{q_k})$ w.r.t.
$\delta_{i_k}^{q_k}$,  a bisection search on $\delta_{i_k}^{q_k}$
can be used to find $\delta_{i_k}^{q_k}(\mu^{q_k})$.  This claim is
established in Appendix \ref{appBisection}. Once
$\delta_{i_k}^{q_k}(\mu^{q_k})$ is found for all
$i_k\in\mathcal{A}_{q_k}$, we can use \eqref{eqVGreaterThanZero} to
find $\bv^{q_k}_{i_k}(\mu^{q_k})$.

{\bf Step 2)} Once
$\{\bv^{q_k}_{i_k}(\mu^{q_k})\}_{i_k\in\mathcal{I}_k}$ is obtained, we need to
search for the optimal $\mu^{q_k}$ that satisfies the feasibility and
the complementarity condition \eqref{eqComplimentarityGeneral}. The following result
(the proof of which can be found in Appendix \ref{appMonotonicityDelta})
suggests that there must exist a $\bar{\mu}^{q_k}>0$ such that the optimal multiplier must lie
in $[0, \bar{\mu}^{q_k}]$. Moreover, we can perform a bisection search to find the optimal multiplier.
\newtheorem{L1}{Lemma}
\begin{L1}\label{lemmaMonotonicityDelta}
{\it For any set of $\{\bv_{i_k}^{q_k}(\mu^{q_k})\}$ that satisfies
\eqref{eqVForceToZero}--\eqref{eqVGreaterThanZero},
$\|\bv_{i_k}^{q_k}(\mu^{q_k})\|$ is strictly decreasing w.r.t.
$\mu^{q_k}$. Moreover, there exists a $\overline{\mu}^{q_k}$ such
that for all $\mu^{q_k}\ge \overline{\mu}^{q_k}$,
$\sum_{i_k\in\mathcal{I}_k}\|\bv^{q_k}_{i_k}(\mu^{q_k})\|^2\le
P_{q_k}$.}
\end{L1}

Performing Step 1) and Step 2) iteratively, we can find the desired optimal
primal-dual pair 
for problem {\rm (P4)}. Table \ref{tableVUpdate} summarizes the above BCD
procedure.

\begin{table*}[htb]
\begin{center}
\vspace{-0.1cm} \caption{ The Procedure for Solving Problem
{\rm (P3)}} \label{tableVUpdate} {\small
\begin{tabular}{|l|}
\hline
\\
S1)\ {\bf Initialization} Generate a feasible set of beamformers $\{\bv^{q_k}_{i_k}\}_{q_k\in\mathcal{Q}_k, i_k\in\mathcal{I}_k}$\\

S2)\ Compute $\bJ_k$ and $\bd_{i_k}$ using \eqref{eqJGeneral} and \eqref{eqDGeneral}\\


S3)\ {\bf Repeat} Cyclically pick a BS $q_k\in\mathcal{Q}_k$\\

S4)\ \quad Compute $\bc_{i_k}$ using \eqref{eqDefC} for each $i_k\in\mathcal{I}_k$ \\
\quad \quad \quad {\bf If} $2\big\|\bc_{i_k}\big\|\le \lambda_k$ set $\bv^{q_k}_{i_k}={\bf 0}$\\

\quad \quad \quad {\bf Else}, choose $\underline{\mu}^{q_k}$ and
$\overline{\mu}^{q_k}$ such that
$(\mu^{q_k})^*\in[\underline{\mu}^{q_k}, ~\overline{\mu}^{q_k}]$\\

S5)\ \quad  \quad {\bf Repeat} $\mu^{q_k}=(\underline{\mu}^{q_k}+\overline{\mu}^{q_k})/2$\\

S6)\ \quad \quad\quad  For each $i_k\in\mathcal{I}_k$, choose
$\underline{\delta}_{i_k}^{q_k}$ and $\overline{\delta}_{i_k}^{q_k}$
such that
$\delta_{i_k}^{q_k}(\mu^{q_k})\in[\underline{\delta}_{i_k}^{q_k},
~\overline{\delta}_{i_k}^{q_k}]$\\

S7)\ \quad\quad\quad  {\bf Repeat} (for each $i_k\in\mathcal{I}_k$)
$\delta_{i_k}^{q_k}=(\underline{\delta}_{i_k}^{q_k}+\overline{\delta}_{i_k}^{q_k})/2$,
\\

S8)\ \quad\quad\quad\quad
$h_{i_k}(\delta_{i_k}^{q_k},\mu^{q_k})=\delta_{i_k}^{q_k}\bigg\|\bigg(\bJ_k[q,q]+(\frac{\lambda_k\delta^{q_k}_{i_k}}{2}
+\mu^{q_k})\bI_M\bigg)^{-1}\bc_{i_k}\bigg\|$\\

S9) \quad\quad\quad\quad  If $h_{i_k}(\delta_{i_k}^{q_k},\mu^{q_k})<1$,
$\underline{\delta}_{i_k}^{q_k}\leftarrow\delta_{i_k}^{q_k}$; Otherwise,

$\overline{\delta}_{i_k}^{q_k}\leftarrow\delta_{i_k}^{q_k}$\\

S10) \quad\quad\quad{\bf Until}
$|\overline{\delta}_{i_k}^{q_k}-\underline{\delta}_{i_k}^{q_k}|\le \epsilon$\\

S11)\quad \quad \quad\ If
$\sum_{i_k\in\mathcal{I}_k}\frac{1}{(\delta^{q_k}_{i_k})^2}<P_{q_k}$,
$\overline{\mu}^{q_k}\leftarrow\mu^{q_k}$; Otherwise,
$\underline{\mu}^{q_k}\leftarrow\mu^{q_k}$ \\

S12) \quad\quad{\bf Until}
$|\overline{\mu}^{q_k}-\underline{\mu}^{q_k}|\le \epsilon$\\

S13) \quad\quad
$\bv^{q_k}_{i_k}\leftarrow\bigg(\bJ_k[q,q]+(\frac{\lambda_k\delta_{i_k}^{q_k}}{2}+\mu^{q_k})\bI_M\bigg)^{-1}
\bc_{i_k}$\\
\quad\quad\quad {\bf End If}\\

S14) {\bf Until} Desired stopping criteria is met\\
\\
  \hline
\end{tabular}}
\vspace{-0.5cm}
\end{center}
\end{table*}

It is worth noting that the algorithm in Table
\ref{tableVUpdate} admits a particular simple form (without
performing the bisection steps) when there is a single user in each
cell, and each BS has a single antenna. Let us again denote the only
user in the $k$th cell as user $k$. The procedure to find BS $q_k$'s
scalar beamformer $v^{q_k}_k\in \mathbb{C}$ for user $k$ (i.e., Step
S4--S13 in Table \ref{tableVUpdate}) can be simplified as follows.
Utilizing \eqref{eqVForceToZero}--\eqref{eqVGreaterThanZero}, if
$|c_k|\le \frac{\lambda_k}{2}$, $v^{q_k}_{k}=0$. Otherwise, note
that $\bJ_k[q,q]\in\mathbb{R}_{+}$ in this case, we have{\small
\begin{align}
&v^{q_k}_k=\frac{c_k}{\bJ_k[q,q]+
\frac{\lambda_k\delta^{q_k}_k}{2}+\mu^{q_k}},
\nonumber\\
&\textrm{with}\quad \delta^{q_k}_k\frac{|c_k|}{\bJ_k[q,q]+
\frac{\lambda_k\delta^{q_k}_k}{2}+\mu^{q_k}}=1.
\end{align}}\hspace{-0.2cm}
Consequently, we obtain a closed-form expression
for $\delta^{q_k}_k$: $\delta^{q_k}_k=\frac{\bJ_k[q,q]+
\mu^{q_k}}{|c_k|-\frac{\lambda_k}{2}}$. Substituting this $\delta^{q_k}_k$ into \eqref{eqVGreaterThanZero},
we obtain{\small
\begin{align}
 v^{q_k}_k=\frac{|c_{k}|-\frac{\lambda_k}{2}}{\bJ_k[q,q]+\mu^{q_k}}
\frac{c_{k}}{|c_{k}|}\nonumber
\end{align}}\hspace{-0.2cm}
where the multiplier $\mu^{q_k}$ should be chosen such that the condition \eqref{eqComplimentarityGeneral}
is satisfied. In summary, we have the following closed-form solution for updating $v^{q_k}_k$
{\small
\begin{align}
v^{q_k}_{k}=\left\{ \begin{array}{ll}
0,&|c_{k}|\le \frac{\lambda_k}{2},\\
\frac{|c_{k}|-\frac{\lambda_k}{2}}{\bJ_k[q,q]}
\frac{c_{k}}{|c_{k}|},&~\textrm{if $\frac{|c_{k}|-\frac{\lambda_k}{2}}{\bJ_k[q,q]}\le \sqrt{P_{q_k}}$},
|c_{k}|> \frac{\lambda_k}{2},\\
\frac{c_k}{|c_k|}\sqrt{P_{q_k}}&~\rm{otherwise.}\label{eqSingleUser}
\end{array}\right.
\end{align}}\hspace{-0.2cm}

The complete algorithm for solving the regularized  weighted MSE
minimization problem ${\rm (P2)}$ is given in Table
\ref{tableWMMSESingle}. We name this algorithm sparse weighted MMSE
algorithm (S-WMMSE). The following theorem states its convergence
property. The proof can be found in Appendix \ref{appConvergence}.

\newtheorem{T1}{\bf Theorem}
\begin{T1}\label{theoremConvergenceVSingle}
{\it The S-WMMSE algorithm converges to a stationary solution of
problem ${\rm (P1)}$.}
\end{T1}

\begin{table}[htb]
\begin{center}
\vspace{-0.1cm} \caption{ The S-WMMSE Algorithm For Solving
Problem ${\rm (P2)}$} \label{tableWMMSESingle} {\small
\begin{tabular}{|l|}
\hline
\\
S1)\ {\bf Initialization} Generate a feasible set of variables $\{\bv_{i_k},  \bu_{i_k},w_{i_k}\}$\\

S2)\ {\bf Repeat} \\

S3)\ \quad $\bu_{i_k}\leftarrow
\left(\sum_{(\ell,j)}\bH^{\ell}_{i_k}\bv_{j_\ell}\bv_{j_\ell}^H(\bH^{\ell}_{i_k})^H
+\sigma^2_{i_k}\bI\right)^{-1}\bH^{k}_{i_k}\bv_{i_k},~\forall~i_k$\\

S4)\ \quad
$w_{i_k}\leftarrow\frac{du_{i_k}(R_{i_k})}{dR_{i_k}}\big|_{R_{i_k}=
R_{i_k}(\bv)}(1-\bu^H_{i_k}\bH^k_{i_k}\bv_{i_k})^{-1}
$, $\forall~i_k$\\

S5)\ \quad For each $k\in\mathcal{K}$, update
$\{\bv^{q_k}\}_{q_k\in\mathcal{Q}_k}$ using Table \ref{tableVUpdate}.\\

S6)\ {\bf Until} Desired stopping criteria is met\\
\\
  \hline
\end{tabular}}
\vspace{-0.3cm}
\end{center}
\end{table}

We remark that in a MISO network in which each user has single
antenna, the algorithm stated in Table
\ref{tableVUpdate}--\ref{tableWMMSESingle} can still be used, except
that in this case the receiver $\bu_{i_k}$ reduces to a scalar.

\subsection{Parameter Selection}\label{subParameter}
In this subsection, we provide guidelines for choosing some key
parameters for practical implementation of the proposed algorithm.

When $\bw,\bu$ are fixed, the procedure in Table
\ref{tableVUpdate} contains two bisection loops for solving for each
$\bv^{q_k}$. The outer loop searches the optimal
$(\mu^{q_k})^*\in[\underline{\mu}^{q_k}, ~\overline{\mu}^{q_k}]$
that ensures $\bv^{q_k}_{i_k}((\mu^{q_k})^*)$ satisfy the
complementarity and feasibility conditions
\eqref{eqComplimentarityGeneral}. The inner loop searches for the
optimal
$\delta^{q_k}_{i_k}(\mu^{q_k})\in[\underline{\delta}^{q_k}_{i_k},~\overline{\delta}^{q_k}_{i_k}]$
to ensure \eqref{eqDeltaEquality}. In implementation, it will be
useful to have explicit
expressions for initial bounds of these variables. 

\subsubsection{The Choice of Initial $\underline{\mu}^{q_k}, ~\overline{\mu}^{q_k}$}
From the fact that $\mu^{q_k}\ge 0$, we can simply set the lower
bound as $\underline{\mu}^{q_k}=0$. For the initial upper
bound $\overline{\mu}^{q_k}$, it is
sufficient to guarantee that
{\small
\begin{align}
\sum_{i_k\in\mathcal{I}_k}\|\bv^{q_k}_{i_k}(\overline{\mu}^{q_k})\|^2<P_{q_k}.\label{eqMuUpperBoundCondition}
\end{align}}\hspace{-0.2cm}
To see this, recall that $\|\bv^{q_k}_{i_k}({\mu}^{q_k})\|^2$ is
monotonically decreasing w.r.t. $\mu^{q_k}$. Consequently when
\eqref{eqMuUpperBoundCondition} is satisfied, there must exist a
$({\mu}^{q_k})^*\in[0,~\overline{\mu}^{q_k}]$ such that both the
feasibility and complementarity condition
\eqref{eqComplimentarityGeneral} is satisfied.

To ensure \eqref{eqMuUpperBoundCondition}, it is sufficient that
such $\overline{\mu}^{q_k}$ satisfies the following condition for
each {\it active user} $i_k\in\mathcal{A}^{q_k}$ (notice that the
active set $\mathcal{A}^{q_k}$ is decided before bisection
starts){\small
\begin{align}
\big\|\bv^{q_k}_{i_k}(\overline{\mu}^{q_k})\big\|=
\big\|\big(\bJ_k[q,q]+(\frac{\lambda_k\delta_{i_k}^{q_k}}{2}+\overline{\mu}^{q_k})\bI_M\big)^{-1}
\bc_{i_k}\big\|<
\sqrt{\frac{P_{q_k}}{{A}^{q_k}}}\nonumber
\end{align}}\hspace{-0.2cm}
For a specific $i_k\in\mathcal{A}^{q_k}$, we have the following
inequalities{\small
\begin{align}
&\bigg\|\bigg(\bJ_k[q,q]+(\frac{\lambda_k\delta_{i_k}^{q_k}}{2}+\overline{\mu}^{q_k})\bI_M\bigg)^{-1}
\bc_{i_k}\bigg\|\label{eqV}\\
&\stackrel{(a)}\le
\rho\left(\bigg(\bJ_k[q,q]+\overline{\mu}^{q_k}\bI_M\bigg)^{-1}\right)\|\bc_{i_k}\|
\stackrel{(b)}\le\frac{1}{\overline{\mu}^{q_k}}\|\bc_{i_k}\|\nonumber
\end{align}}\hspace{-0.2cm}
where in $(a)$ we have used the fact
that \eqref{eqV} is decreasing w.r.t. $\delta^{q_k}_{i_k}$; 
in $(b)$ we have used the fact that
$\bJ_k[q,q]\succeq 0$. As a result, it is sufficient to find a
$\overline{\mu}^{q_k}$ such that $
\frac{1}{\overline{\mu}^{q_k}}\|\bc_{i_k}\|<\sqrt{\frac{P_{q_k}}{{A}^{q_k}}}$ for all $i_k\in\mathcal{A}^{q_k}$.
This implies that the following choice ensures \eqref{eqMuUpperBoundCondition} 
{\small
\begin{align}
\overline{\mu}^{q_k}>\left({\frac{P_{q_k}}{{A}^{q_k}}}\right)^{-\frac{1}{2}}\max_{i_k\in\mathcal{A}^{q_k}}\|\bc_{i_k}\|\label{eqMuUpperBound}.
\end{align}}\hspace{-0.2cm}
\vspace{-0.3cm}
\subsubsection{The Choice of Initial $\underline{\delta}^{q_k}_{i_k}$, $\overline{\delta}^{q_k}_{i_k}$}
Once the initial bounds on ${\mu}^{q_k}$ are chosen, we can
determine the initial bounds for each $\delta^{q_k}_{i_k}$,
$i_k\in\mathcal{A}^{q_k}$. Because $\delta^{q_k}_{i_k}>0$, the lower
bound can be simply set as $\underline{\delta}^{q_k}_{i_k}=0$. Next,
we will find the $\overline{\delta}^{q_k}_{i_k}$ that is suitable
for all $\mu^{q_k}\in[\underline{\mu}^{q_k}, \overline{\mu}^{q_k}]$.
From the proof of Lemma \ref{lemmaExistenceDelta}, we see that it is
sufficient to choose the initial bound $\overline{\delta}^{q_k}_{i_k}$ such that{\small
\begin{align}
h_{i_k}(\overline{\delta}^{q_k}_{i_k},\mu^{q_k})>1, \
\forall~\mu^{q_k}\in[\underline{\mu}^{q_k},
\overline{\mu}^{q_k}], \forall~i_k\in\mathcal{A}^{q_k}\label{eqCriteriaChoosingDelta}.
\end{align}}\hspace{-0.2cm}
In this way for each $\mu^{q_k}\in[\underline{\mu}^{q_k},
\overline{\mu}^{q_k}]$ there is a
${\delta}^{q_k}_{i_k}(\mu^{q_k})\in[0,
\overline{\delta}^{q_k}_{i_k}]$ that ensures {\small
$h_{i_k}({\delta}^{q_k}_{i_k}(\mu^{q_k}),\mu^{q_k})=1$}.

We have the following series of inequalities bounding
$h_{i_k}(\overline{\delta}^{q_k}_{i_k},\mu^{q_k})$ for any $\mu^{q_k}\in[\underline{\mu}^{q_k},
\overline{\mu}^{q_k}]${\small
\begin{align}
h_{i_k}(\overline{\delta}^{q_k}_{i_k},\mu^{q_k})&
\ge\overline\delta_{i_k}^{q_k}\bigg\|\bigg(\bJ_k[q,q]+(\frac{\lambda_k\overline{\delta}^{q_k}_{i_k}}{2}
+\overline{\mu}^{q_k})\bI_M\bigg)^{-1}\bc_{i_k}\bigg\|\nonumber\\
&\ge\overline{\delta}_{i_k}^{q_k}\frac{1}{\rho(\bJ_k[q,q])+\frac{\lambda_k\overline{\delta}^{q_k}_{i_k}}{2}
+\overline{\mu}^{q_k}}\big\|\bc_{i_k}\big\|\label{eqHLowerBounds}
\end{align}}\hspace{-0.2cm}
where the first inequality is due to the monotonicity of
$h_{i_k}(\overline{\delta}^{q_k}_{i_k},\mu^{q_k})$ w.r.t.
$\mu^{q_k}$.
Using \eqref{eqHLowerBounds},
we can show that the following choice is a sufficient condition for \eqref{eqCriteriaChoosingDelta}{\small
\begin{align}
\overline{\delta}^{q_k}_{i_k}>\frac{1}{\big\|\bc_{i_k}\big\|-\frac{\lambda_k}{2}}(\rho(\bJ_k[q,q])+\overline{\mu}^{q_k}),
~\forall~i_k\in\mathcal{A}^{q_k}\nonumber.
\end{align}}\hspace{-0.2cm}
\vspace{-0.5cm}
\subsection{Distributed Implementation}\label{subImplementation}
Suppose that there is some central entity, say a macro BS,
managing the downlink resource allocation for each cell. Then under the
following assumptions, the proposed algorithm can be implemented distributedly by each macro BS.
\begin{description}
\item [{\bf A-1)}] each macro BS $k$ knows the channels from the BSs in its cell
to all the users $\mathcal{I}$;

\item [{\bf A-2)}] each user
has an additional channel to feedback information to its current serving BS;

\item [{\bf A-3)}] different macro BSs
can exchange control information.
\end{description}
Under these assumptions, in each iteration of the algorithm, a user
$i_k$ can measure the covariance of the received signal $\bC_{i_k}$
and update its weight and receive beamformer $w_{i_k}, \bu_{i_k}$,
respectively. It then feeds these variables to one of its serving
BS, who in turn forwards it to the macro BS. Each pair of macro BSs
then exchange their respective users' current beamformers. With
these pieces of information, all macro BSs can carry out the
procedure in Table \ref{tableVUpdate} independently. The newly
computed beamforming and clustering decisions are subsequently
distributed to the BSs in their respective cells via low-latency
backhaul links.

In practice, considering the costs of obtaining and sharing of the
channel state information, the sparse clustering algorithm may only need
to be executed in its full generality in every several transmission
time intervals (TTIs). During the TTIs in which the clustering is
kept fixed, one can either keep updating the beamformers (by solving
problem (P1) {\it without} the regularization term), or even fix
the beamformers.

\section{Numerical Results}\label{secSimulation}
\begin{figure}[htb] \vspace*{-.3cm}
    {\includegraphics[width=
    0.6 \linewidth]{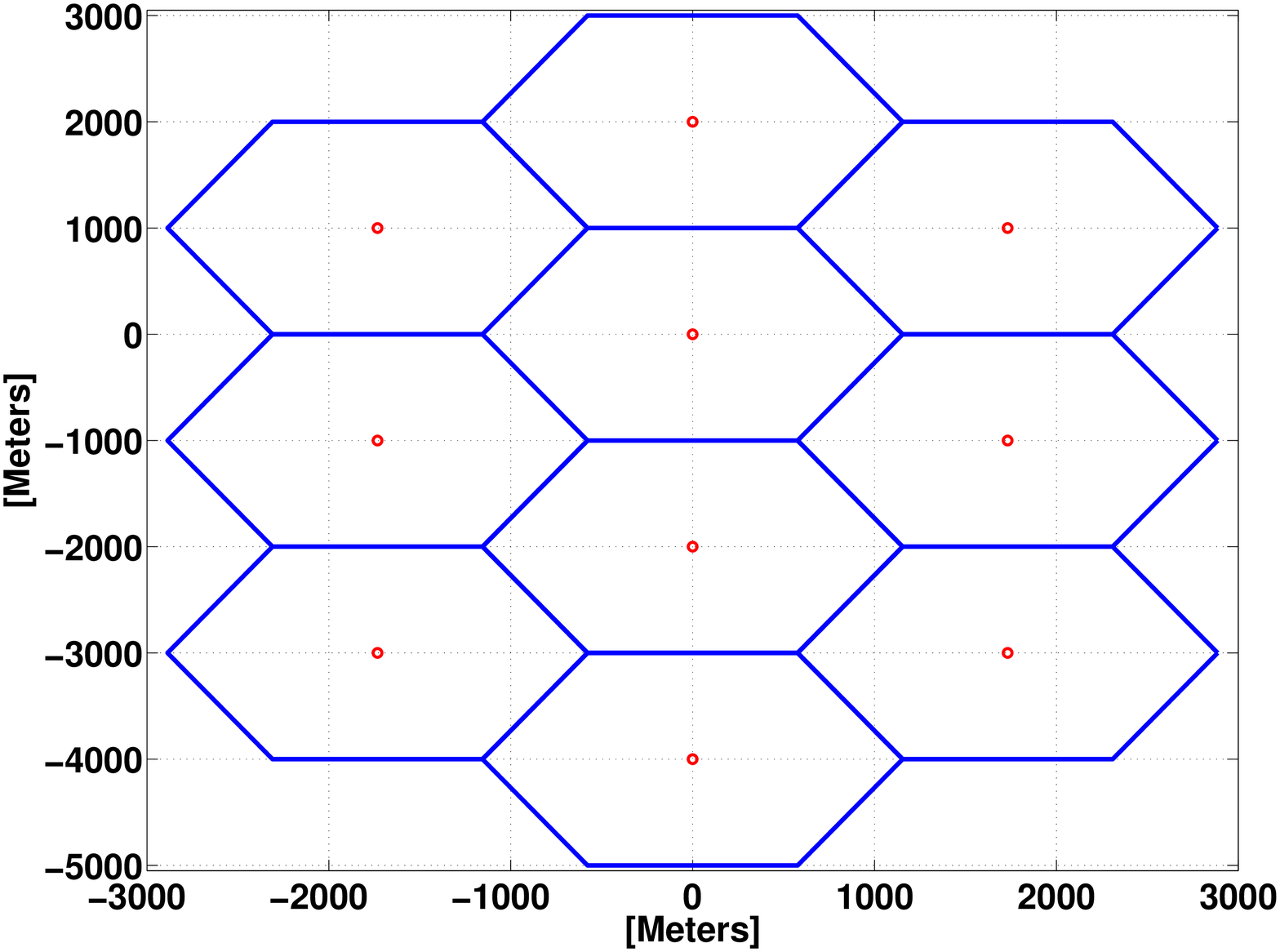}
    \vspace*{-0.2cm}\caption{ Cell configuration for numerical experiments. }\label{figCell}
    \vspace*{-0.1cm}}
    \end{figure}

In this section, we perform numerical evaluation of our proposed
algorithm. We consider a multicell network of up to $10$ cells. The
distance of the centers of two adjacent cells is set to be $2000$
meters (see Fig. \ref{figCell} for an illustration). We place the
BSs and users randomly in each cell. Let $d^{q_\ell}_{i_k}$ denote
the distance between BS $q_\ell$ and user $i_k$. The channel
coefficients between user $i_k$ and BS $q_\ell$ are modeled as zero
mean circularly symmetric complex Gaussian vector with
$\left({200}/{d^{q_\ell}_{i_k}}\right)^{3}L^{q_\ell}_{i_k}$ as
variance for both real and imaginary dimensions, where
$10\log10(L^{q_\ell}_{i_k})\sim\mathcal{N}(0,64)$ is a real Gaussian
random variable modeling the shadowing effect. We fix the
environment noise power for all the users as $\sigma^2_{i_k}=1$, fix
the power budget of each BSs as $P_{q_k}=P$, and fix the number of
BSs and the number of users in each cell as $|\mathcal{Q}_k|=Q$,
$|\mathcal{I}_k|=I$. We define ${\rm SNR}={P}Q$.

\begin{figure*}[htb] \vspace*{-.3cm}
    \begin{minipage}[t]{0.48\linewidth}
    \centering
    {\includegraphics[width=
1  \linewidth]{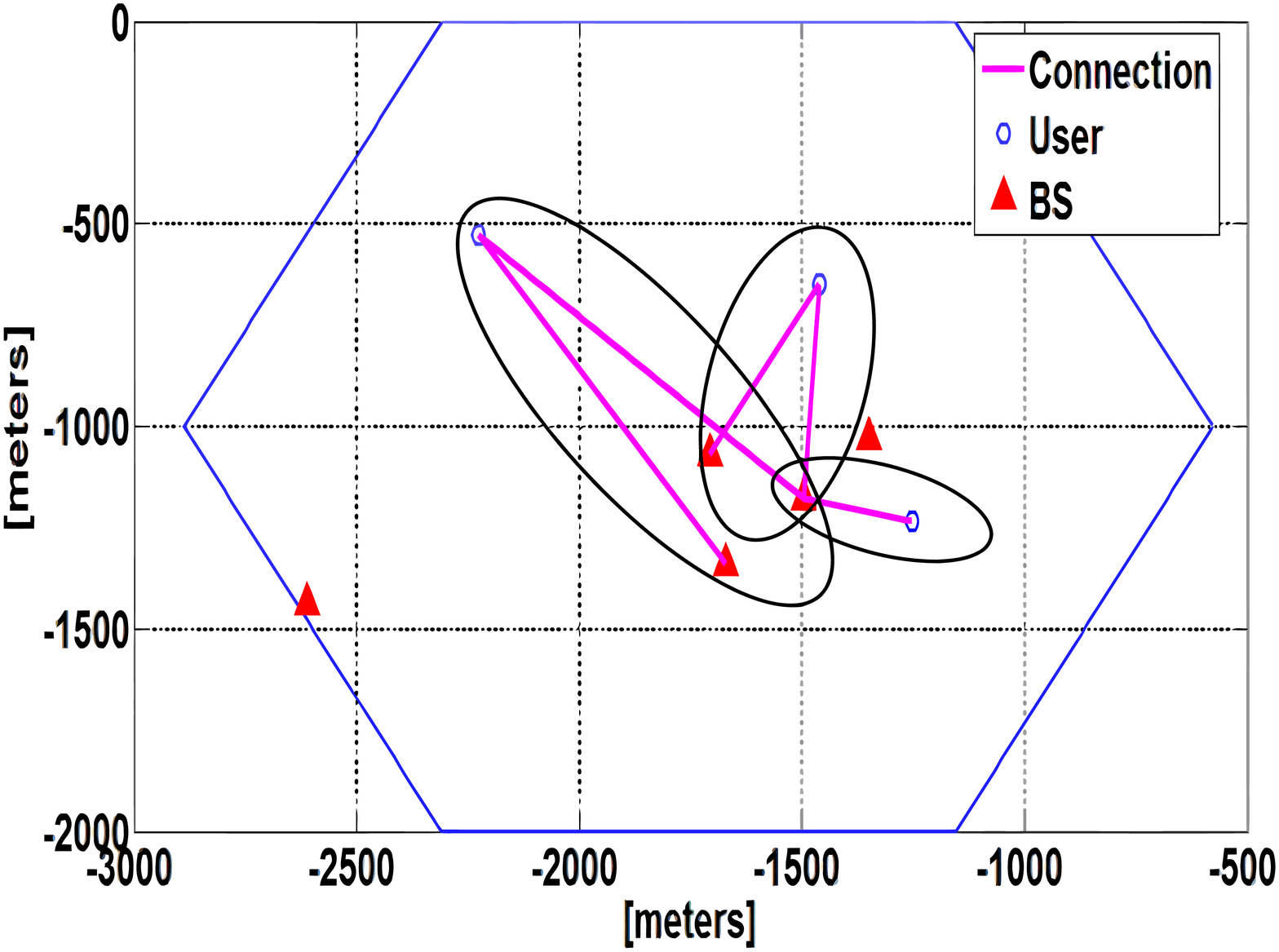}
\vspace*{-0.2cm}\caption{Illustration of coordination
    status of running the S-WMMSE algorithm for a single cell network. A line connecting a BS and
    a user means this BS currently serves this user. The black
    circles indicate the coordination clusters generated by the algorithm.
    $K=1$, $M=4$, $N=2$, $|\cI_1|=3$, $|\cQ_1|=5$. The sum rate utility is used.}\label{figConfigurationSparse}
    \vspace*{-0.1cm}}
\end{minipage}\hfill
    \begin{minipage}[t]{0.48\linewidth}
    \centering
    {\includegraphics[width=
1  \linewidth]{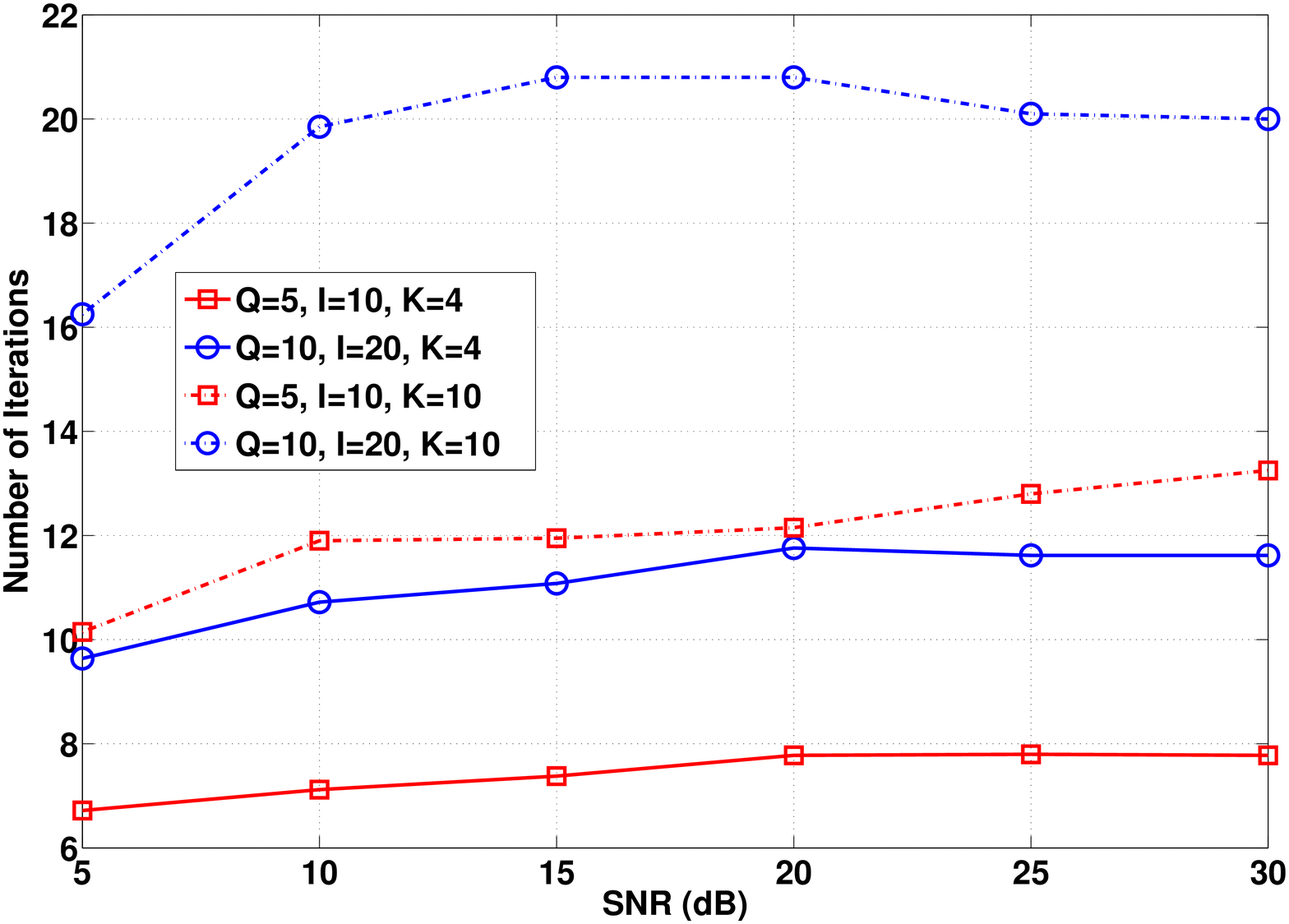}
\vspace*{-0.2cm}\caption{Comparison of the number of iterations
needed for convergence with different network sizes. $K=\{4,10\}$,
$M=4$, $N=2$,  $\lambda_k=\frac{QK}{I\sqrt{\mbox{SNR}}}, \ \forall \
k$. The sum rate utility is used. }\label{figIterations}
\vspace*{-0.1cm}}
\end{minipage}
\vspace*{-0.1cm}
    \end{figure*}

In Fig. \ref{figConfigurationSparse}, we illustrate the structure of
the overlapping clusters generated by the S-WMMSE algorithm in a
simple single cell network. In this example, when no group-sparsity
is considered, each user is served by all the BSs in the set
$\mathcal{Q}_1$. In contrast, when the clusters are formed by
performing the proposed algorithm, the cluster sizes are
significantly reduced. In Fig. \ref{figIterations}, we show the
averaged number of iterations\footnote{Note that the number of
iterations refer to the {\it outer iterations} specified by the
update in Table \ref{tableWMMSESingle}. } needed for the proposed
algorithm to reach convergence in different network scenarios. The
stopping criteria is set as $|f(\bv^{t+1})-f(\bv^{t}))|<10^{-1}$,
where $f(\cdot)$ represents the objective value of problem (P1). For
both of these results, the sum rate utility is used.

Our experiments mainly compare the proposed algorithm with the following
three algorithms:
\begin{itemize}

\item {\it WMMSE with full intra-cell and limited inter-cell coordination \cite{shi11WMMSE_TSP}}:
In this algorithm, the network is modeled as a MIMO-IBC where
all the BSs $\mathcal{Q}_k$ in cell $k$ collectively form a giant virtual BS with the transmit power pooled together. It is shown in \cite{shi11WMMSE_TSP} that this algorithm compares favorably to other
popular beamformer design algorithms such as the iterative pricing algorithm \cite{kim11}. In
the present paper, it corresponds to the case where a single
cluster is formed in each cell. This algorithm serves as a performance upperbound (in terms of throughput)
of the proposed algorithm.

\item {\it ZF beamforming with heuristic BS clustering}:
In this algorithm, cooperation clusters with fixed sizes are formed within each BS, and each cluster
performs ZF beamforming. The clusters are formed greedily by
choosing an initial BS among the unclustered BSs and adding its nearest BSs until reaching the
prescribed cluster size. The users are assigned to the cluster with the strongest direct channel (in terms of
2-norm). Each cluster serves its associated users by a single
cell ZF linear beamforming \cite{Spencer04} \footnote{Note that in \cite{Spencer04}, after the
beams are calculated and fixed, the power allocation for different beams/streams are determined by
solving a convex vector optimization with sum-power constraint. In our simulation, we replace the sum-power
constraint with a set of  per-group of antenna power constraint to better fit the multi-BS setting. }. To ensure
feasibility of the per-cluster ZF scheme, the weakest users in terms of direct channel
are dropped when infeasibility arises.

\item {\it WMMSE with each user served by its nearest BS}: In this algorithm, each user is assigned to
the nearest BS,  that is, the size of the coordination cluster is at most $1$.
We denote this algorithm as WMMSE-nearest neighbor (NN).
\end{itemize}

 \begin{figure*}[htb] \vspace*{-.1cm}
    \begin{minipage}[t]{0.48\linewidth}
    \centering
    {\includegraphics[width=
    1 \linewidth]{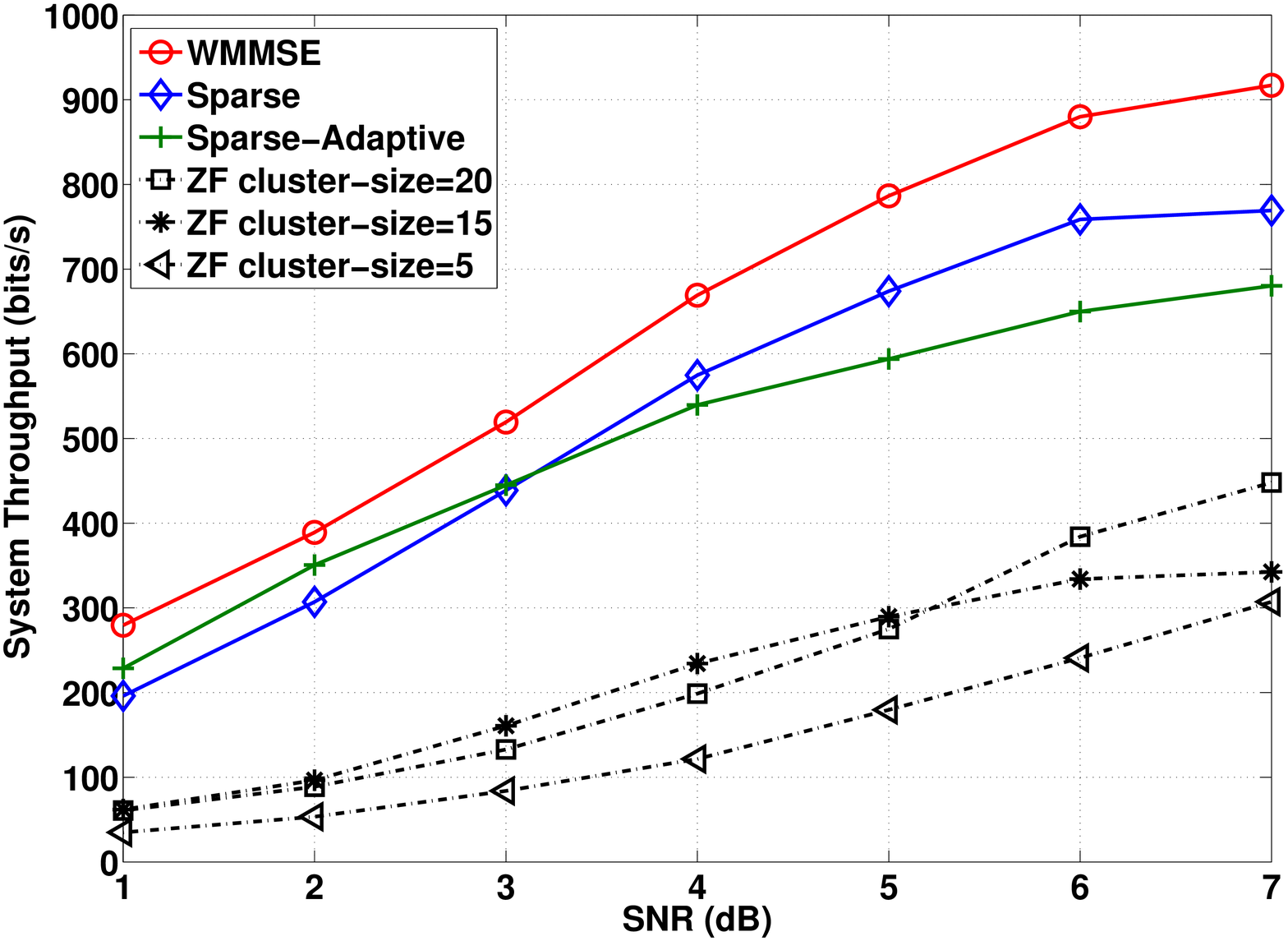}
    \vspace*{-0.2cm}\caption{ Comparison of the system throughput achieved by different algorithms.
    $K=4$, $M=4$, $N=2$, $|\cI_k|=40$, $|\cQ_k|=20$, the sum rate utility is used. For the S-WMMSE algorithm,
    $\lambda_k=\frac{QK}{I\sqrt{\mbox{SNR}}}, \ \forall \ k$.}\label{figRateCompareZF}
    \vspace*{-0.1cm}}
\end{minipage}\hfill
\begin{minipage}[t]{0.48\linewidth}
    \centering
    {\includegraphics[width=
    1 \linewidth]{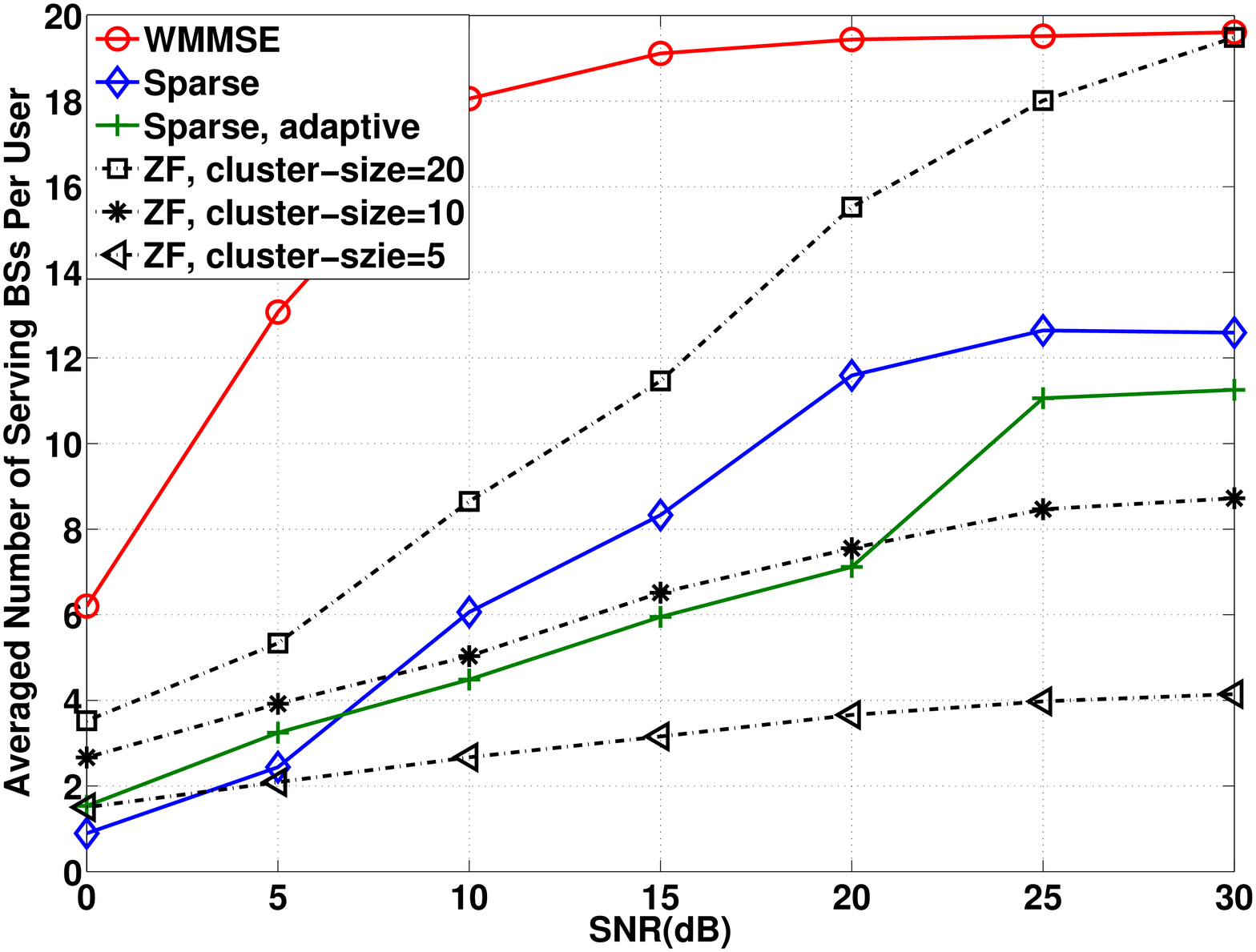}
    \vspace*{-0.2cm}\caption{ Comparison of the averaged number of BSs serving each user for different algorithms.
    $K=4$, $M=4$, $N=2$, $|\cI_k|=40$, $|\cQ_k|=20$, the sum rate utility is used. For the S-WMMSE algorithm,
    $\lambda_k=\frac{QK}{I\sqrt{\mbox{SNR}}}, \ \forall \ k$.}\label{figSparsityCompareZF}
    \vspace*{-0.1cm}}
\end{minipage}
\vspace*{-0.1cm}
    \end{figure*}

We first consider a network with $K=4$, $I=40$, $Q=20$, $M=4$, $N=2$. We use system sum rate
as the utility function. The achieved system sum rate
and the averaged number of serving BSs are shown in Fig. \ref{figRateCompareZF}-- Fig. \ref{figSparsityCompareZF}
for different algorithms.
Each point in the figures is an average of $100$ runs of the
algorithms over randomly generated networks. Notice that for the ZF
based scheme, although the cluster size is given and fixed, the
actual number of serving BSs per user can be smaller than the
cluster size, as some users may not be served by all the BSs in its
serving cluster. It can be seen from Fig. \ref{figRateCompareZF}--
Fig. \ref{figSparsityCompareZF} that the system throughput obtained
by the proposed S-WMMSE algorithm is close to what is achievable by
the full cooperation. Moreover, the high throughput is achieved
using moderate cluster sizes.  Notice that the proposed algorithm
compares favorably even with the full per-cell ZF scheme (with
cluster size $20$). This suggests that the inter-cluster
interference should be carefully taken into consideration when
jointly optimizing the BS clusters and beamformers.

It is important to emphasize that the parameters
$\{\lambda_k\}_{k=1}^{K}$ in the proposed algorithm balance the
sizes of the clusters and the system throughput. For different
network configurations they need to be properly chosen to yield the
best tradeoff. Empirically, we found that setting
$\lambda_k=\frac{QK}{I\sqrt{\mbox{SNR}}}$ gives a satisfactory
tradeoff (as illustrated in Fig. \ref{figRateCompareZF}-- Fig.
\ref{figSparsityCompareZF}). This is partly because choosing
$\lambda_k$ inversely proportional to $\sqrt{\mbox{SNR}}$ can better
balance the relative importance of the penalization term and the sum
rate term when $\mbox{SNR}$ becomes large. To better select this
parameter for different system settings, we provide an alternative
scheme that adaptively computes $\{\lambda_k\}_{k=1}^{K}$ {\it in
each iteration} of the algorithm. Note that for the quadratic
problem {\rm (P4)}, if $\lambda_k$ is chosen large enough such that
$\lambda_k>\bar{\lambda}_k\triangleq
2\times\max_{q\in\mathcal{Q}_k,i_k\in\mathcal{I}_k}\|\bd_{i_k}[q]\|$,
then $\bv_{i_k}={\bf{0}}, ~\forall~i_k\in\mathcal{I}_k$. This result
can be straightforwardly derived from the optimality condition
\eqref{eqVForceToZero}. For conventional quadratic LASSO problem,
the fixed sparsity parameter $\lambda_k$ can be chosen as
$c\bar{\lambda}_k$, where $0<c<1$ is a small number, see, e.g.,
\cite{Wright09}. In our experiments, we found that choosing
$\lambda_k$ as $\lambda_k=\min\big\{\frac{0.01\bar{\lambda}_k}{{\rm
SNR}},1\big\}$ works well for all network configurations. The
performance of the S-WMMSE algorithm with this adaptive choices of
$\lambda_k$ is also demonstrated in Fig.
\ref{figRateCompareZF}--Fig. \ref{figSparsityCompareZF}. Clearly
such adaptive choice of $\lambda_k$ can generate smaller sizes of
the clusters while achieving similar performance as its fixed
parameter counterparts. Note that the convergence proof for the
proposed algorithm does not apply anymore, as it requires that
parameters $\{\lambda_k\}$ must be fixed during the iterations
(although in simulation experiments we observe that this adaptive
algorithm usually converges).

 \begin{figure*}[htb] \vspace*{-.1cm}
    \begin{minipage}[t]{0.48\linewidth}
    \centering
    {\includegraphics[width=
    1 \linewidth]{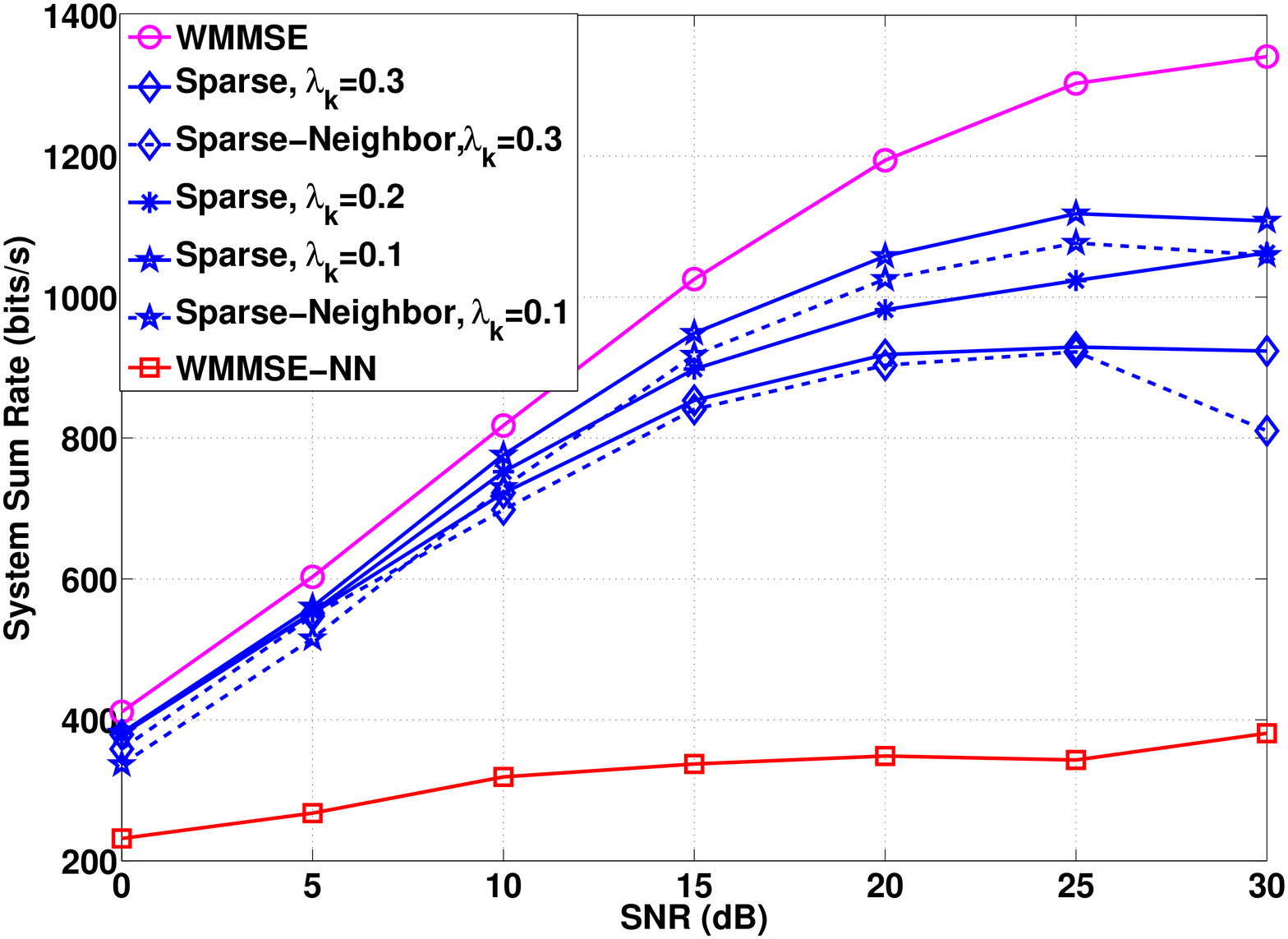}
    \vspace*{-0.1cm}\caption{Comparison of the system throughput achieved by different algorithms.
    $K=10$, $M=4$, $N=2$, $|\cI_k|=20$, $|\cQ_k|=20$, the PF utility is used. $\lambda_k$ is specified in the legend.}\label{figRatePF}
    \vspace*{-0.1cm}}
\end{minipage}\hfill
    \begin{minipage}[t]{0.48\linewidth}
    \centering
    {\includegraphics[width=
1  \linewidth]{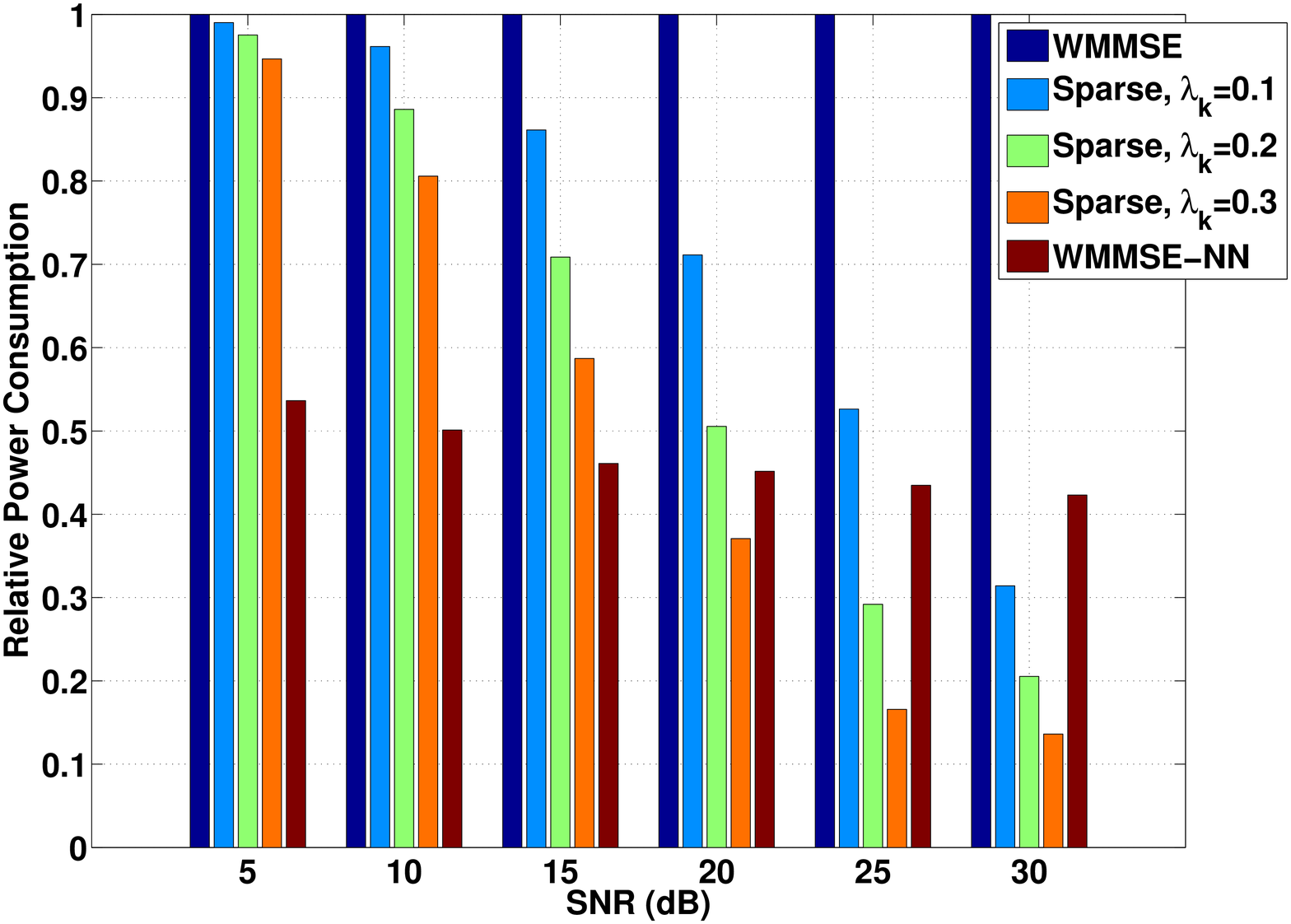} \vspace*{-0.1cm}\caption{Comparison
of relative per-BS transmission power used (relative to the power
consumption of WMMSE algorithm with full per-cell cooperation).
    $K=10$, $M=4$, $N=2$, $|\cI_k|=20$, $|\cQ_k|=20$, PF utility is used. $\lambda_k$ is specified in the legend.}\label{figPowerPF}
\vspace*{-0.1cm}}
\end{minipage}
\vspace*{-0.1cm}
    \end{figure*}

         \begin{figure*}[htb] \vspace*{-.1cm}
    \begin{minipage}[t]{0.48\linewidth}
    \centering
    {\includegraphics[width=
    1 \linewidth]{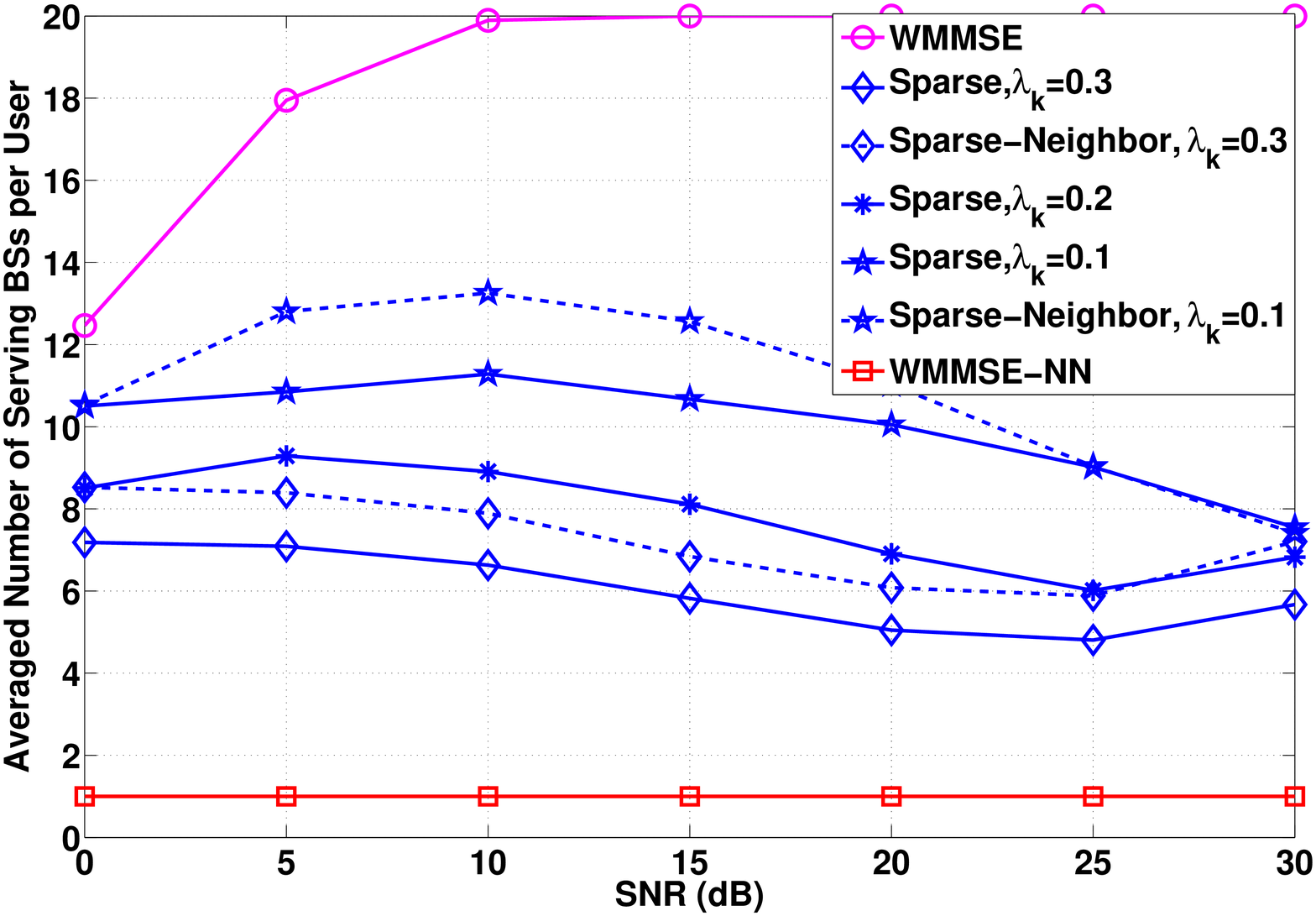}
    \vspace*{-0.2cm}\caption{Comparison of the averaged cluster sizes generated by different algorithms.
    $K=10$, $M=4$, $N=2$, $|\cI_k|=20$, $|\cQ_k|=20$, PF utility is used.  $\lambda_k$ is specified
in the legend. }\label{figSparsityPF}
    \vspace*{-0.1cm}}
\end{minipage}\hfill
    \begin{minipage}[t]{0.48\linewidth}
    \centering
    {\includegraphics[width=
1  \linewidth]{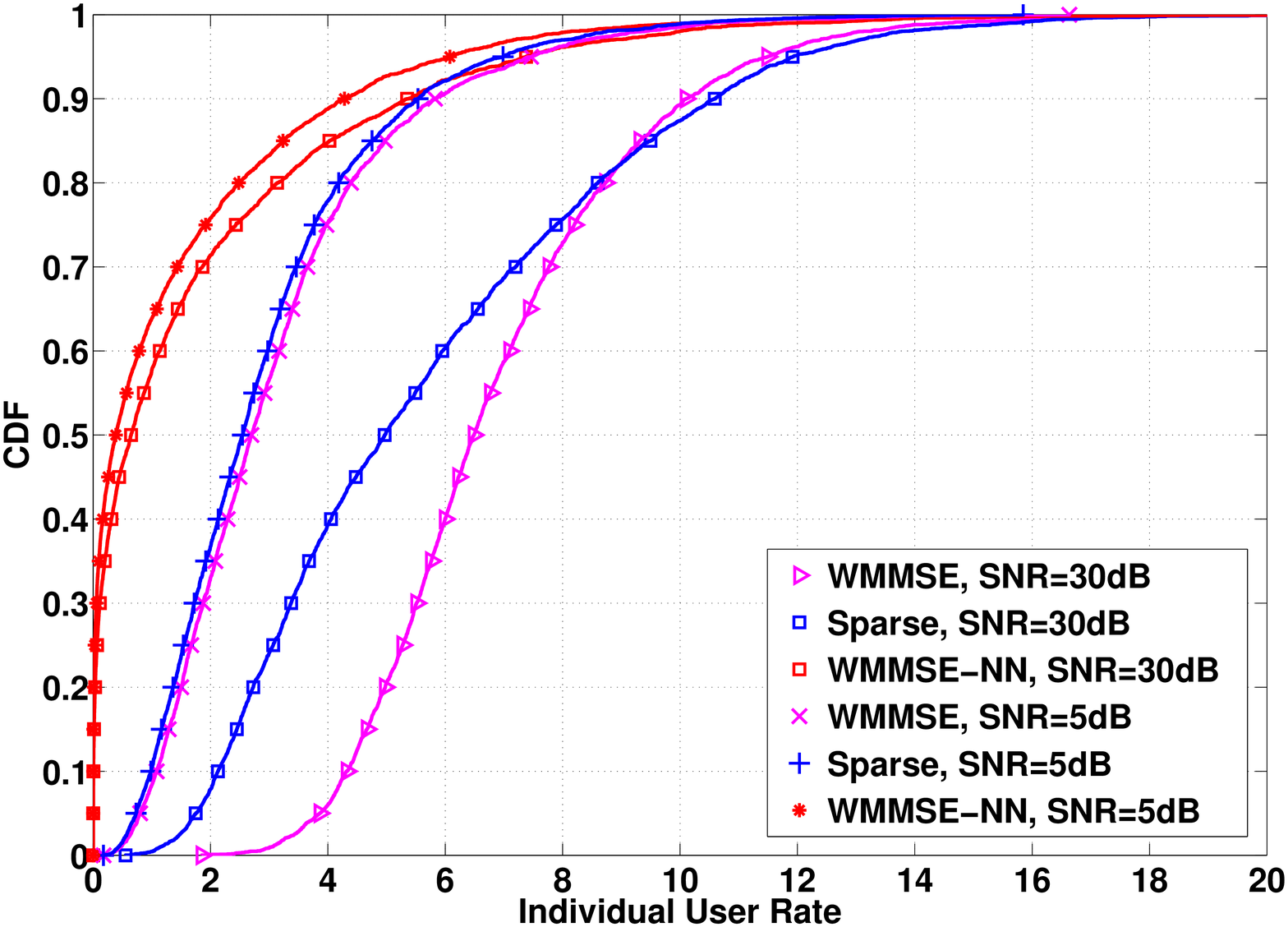} \vspace*{-0.2cm}\caption{Comparison
of distribution of the users' individual transmission rates achieved
by different algorithms. $K=10$, $M=4$, $N=2$, $|\cI_k|=20$,
$|\cQ_k|=20$, PF utility is used. For the S-WMMSE algorithm,
$\lambda_k=0.1$.}\label{figCDFPF} \vspace*{-0.1cm}}
\end{minipage}
\vspace*{-0.4cm}
    \end{figure*}

We also consider a larger network with $K=10$, $I=Q=20$, $M=4$ and
$N=2$, and choose to optimize the proportional fairness utility
defined as $u_{i_k}(R_{i_k})=\log(R_{i_k})$. In Fig.
\ref{figRatePF}--Fig. \ref{figCDFPF}, we compare the performance of
the WMMSE algorithm and the WMMSE-NN algorithm\footnote{We do not
consider the ZF scheme in this experiment for the reason that it
cannot guarantee that all the users in the system are served
simultaneously, as required by the solution of the proportional fair
utility maximization problem.} with the proposed algorithm for
different choices of $\lambda_k$. In order to highlight the role of
$\lambda_k$ in balancing the system throughput and the cluster
sizes, we show in these figures the performance of the proposed
algorithm with {\it fixed} sparsity parameter $\lambda_k$ { for all
SNR values}. In Fig.\ \ref{figPowerPF}, we plot the averaged per-BS
power consumption relative to that of the WMMSE algorithm. In Fig.\
\ref{figCDFPF} we plot the distribution of the individual users'
rates generated by these algorithms. Clearly the proposed algorithm
is able to achieve high levels of system throughput and fairness by
only using small cluster sizes and significantly lower transmission
power (the reduction of transmission power can also be attributed to
the use of penalization, see Fig. \ \ref{figPowerPF}).
 Additionally, in Fig.\ \ref{figRatePF} and Fig.\
\ref{figSparsityPF}, we include the performance of a limited
cooperation scheme in which each cell only coordinates with its
nearest neighbor, while treating the signals of the remaining cells
as thermal noise (this scheme is labeled as ``Sparse-Neighbor"). We
observed that this scheme has similar system throughput as the
original one, but results in larger cluster size. Such increase in
cluster size can be seen as a compensation adopted by the
``Sparse-Neighbor" algorithm for ignoring certain inter-cell
interference. Due to the limited coordination among the cells, the
convergence of this scheme is not theoretically guaranteed. However
in simulation we found that the algorithm usually converges.

\section{Concluding Remarks}\label{secConclusion}
In this work, we propose to jointly optimize the BS clustering and downlink linear beamformer in a large
scale HetNet by solving a nonsmooth utility maximization problem. A key observation that motivates this work
 is that when all the BSs in each cell is viewed as a single virtual BS, the limited coordination
strategy that requires a few BSs jointly transmit to a user is equivalent to a group-sparsity
structure of the virtual BSs' beamformers. We effectively incorporate such group-sparsity
into our beamformer design by penalizing the system utility function using a mixed $\ell_2/\ell_1$ norm.
We derive a useful equivalent reformulation of this nonsmooth utility maximization problem, which
facilitates the design of an efficient iterative group-LASSO based algorithm. Simulation results
show that the proposed algorithm is able to select a few serving BSs for each user, while incurring
minor loss in terms of system throughput and/or user fairness.

Our framework can be extended for the scenario that multiple streams
are transmitted to each user as well. In this more general case, a
precoding matrix is used by each BS for each user. To induce
sparsity, the utility function should be penalized by the Frobenius
norms of the precoding matrices. All the equivalence results derived
in Section \ref{secEquivalent} hold true for this general case,
while the algorithm needs to be properly tailored. We also expect
that the proposed approach can be extended for other related
problems such as the design of coordinated transceiver in an uplink
HetNet, or the design of antenna selection algorithms for large
scale distributed antenna systems.

\section{Acknowledgement}
We wish to thank the anonymous reviewers for their comments that
improve the presentation of the paper. The first author also wishes
to thank Dr. Qiang Li from Chinese University of Hong Kong and Dr.
Ya-feng Liu from Chinese Academy of Sciences for many helpful
discussions.

\appendices
\section{Proof of Proposition \ref{propEquivalence} and Proposition \ref{propEquivalenceGeneral}}\label{appEquivalence}
{\it Proof of Proposition \ref{propEquivalence}}: Let
$I(\bv_k^{q_k})$ denote the (nonsmooth) indicator function for the
feasible space of vector $\bv^{q_k}_k$, i.e.,{\small
\begin{align}
I(\bv_k^{q_k})&= \left\{ \begin{array}{ll}
1,&(\bv^{q_k}_{k})^H\bv^{q_k}_{k}\le P_{q_k},\nonumber\\
-\infty,&\textrm{otherwise}.
\end{array}\right.
\end{align}}\hspace{-0.1cm}
We can then rewrite problem
\eqref{problemSingle} compactly as
an unconstrained nonsmooth optimization problem{\small
\begin{align}
&\max_{\{\bv^{q_k}_{k}\}}\sum_{k\in\mathcal{K}}\bigg(R_{k}
-\lambda_k\sum_{q_k\in\mathcal{Q}_k}\|\bv^{q_k}_{k}\|+\sum_{q_k\in\mathcal{Q}_k}I(\bv^{q_k}_k)\bigg)
\triangleq \max_{\{\bv^{q_k}_{k}\}} f_{\rm
R}(\bv)\label{problemSingleUnconstrained}
\end{align}}\hspace{-0.1cm}
and rewrite \eqref{problemWMMSESingle} equivalently as{\small
\begin{align}
&\min_{\{\bv^{q_k}_{k},\bu_k,w_k\}}\sum_{k\in\mathcal{K}}\bigg(w_k
e_k-\log(w_k)+
\lambda_k\hspace{-0.2cm}\sum_{q_k\in\mathcal{Q}_k}\|\bv^{q_k}_k\|-\hspace{-0.2cm}\sum_{q_k\in\mathcal{Q}_k}I(\bv^{q_k}_k)\bigg)\nonumber\\
&\triangleq \min_{\{\bv^{q_k}_{k}\}, \{\bu_k\},
\{w_k\}}f_{\rm{mse}}(\bv,\bu,\bw)\label{problemWMMSESingleUnconstrained}.
\end{align}}\hspace{-0.1cm}
We first claim that the function $f_R(\cdot)$ is {\it regular} in
the sense of Definition \ref{defRegular}, under the block structure
$\bv=\{\bv_k^{q_k}\}_{q_k\in\mathcal{Q}_k, k\in\mathcal{K}}$. This
claim can be verified by observing that under such block structure,
the nonsmooth parts of the function
$f_R(\cdot)$ 
are {\it separable} across blocks, and the smooth part of
$f_R(\cdot)$ is differentiable. This property ensures that  a
coordinatewise stationary point $\bv$ of $f_R(\cdot)$ is also a
stationary point. See \cite[Lemma 3.1]{tseng01} for a derivation.
Similarly, the function $f_{\rm{mse}}(\cdot)$ is regular under the
following block structure{\small
\begin{align}
\bv=\{\bv_k^{q_k}\}_{q_k\in\mathcal{Q}_k, k\in\mathcal{K}},\
\bu=\{\bu_k\}_{k\in\mathcal{K}},\ \bw=\{w_k\}_{k\in\mathcal{K}}.\nonumber
\end{align}}\hspace{-0.2cm}
Now assume that $(\bv^*,\bu^*, \bw^*)$ is a stationary solution of
problem \eqref{problemWMMSESingleUnconstrained}, then we must
have{\small
\begin{align}
&f_{\rm{mse}}'(\bv^*,\bu^*,\bw^*;
(0,\cdots,\bd_{\bu_k},0,\cdots,0))\ge 0, \ \forall~\bd_{\bu_k},\
\forall~k\label{eqUCondition}\\
&f_{\rm{mse}}'(\bv^*,\bu^*,\bw^*; (0,\cdots,d_{w_k},0,\cdots,0))\ge
0, \ \forall~d_{w_k}, \ \forall~k\label{eqWCondition}
\end{align}}\hspace{-0.2cm}
where $(0,\cdots,\bd_{\bu_k},0,\cdots,0)$ is a vector of zero entries except for
the block corresponding to the variable $\bu_k$, which takes the value $\bd_{\bu_k}$.
Condition \eqref{eqUCondition} implies that $\bu^*_k$ is the
unconstrained local minimum of the function
$f_{\rm{mse}}(\bv^*,[\bu_k, \bu^*_{-k}],\bw^*)$. The same is true for  $w_k$. Notice that $f_{\rm{mse}}(\cdot)$ is smooth w.r.t. $\bu_k$ and
$w_k$, consequently, we must have{\small
\begin{align}
\triangledown_{\bu_k}f_{\rm{mse}}(\bv^*,\bu^*,\bw^*)=0,\
\frac{\partial f_{\rm{mse}}(\bv^*,\bu^*,\bw^*)}{\partial w_k}=0,\ \forall~
k\in\mathcal{K}\nonumber.
\end{align}}\hspace{-0.2cm}
The above two sets of conditions imply that{\small
\begin{align}
\bu^*_k&=\bC^{-1}_k(\bv^*)\bH^k_k\bv^*_k,\nonumber\\
w^*_k&=\frac{1}{e^*_k}=\left(1-(\bv^*_k)^H(\bH^k_k)^H
\bC^{-1}_k(\bv^*)\bH^k_k\bv^*_k\right)^{-1}.\nonumber
\end{align}}\hspace{-0.2cm}
In the sequel we will occasionally use $\bu^*_k(\bv^*)$ and
$w^*_k(\bv^*)$ to emphasize their dependencies on $\bv^*$. Using
these two expressions, we have{\small
\begin{align}
&f_{\rm{mse}}(\bv^*,\bu^*,\bw^*)\nonumber\\
&=\sum_{k\in\mathcal{K}}\bigg(1-\log((e^*_k)^{-1})+
\lambda_k\sum_{q_k\in\mathcal{Q}_k}\|\bv^{q_k}_k\|-\sum_{q\in\mathcal{Q}_k}I(\bv^{q_k}_k)\bigg)\nonumber\\
&\stackrel{(a)}=K-f_{R}(\bv^*)\label{eqEquivalenceMMSERate}
\end{align}}\hspace{-0.2cm}
where in $(a)$ we have used the matrix inversion lemma
\cite{horn90} to obtain{\small
\begin{align}
&\log((e^*_k)^{-1})=\log\bigg(\bigg(1-(\bv^*_k)^H(\bH^k_k)^H
\bC_k^{-1}(\bv^*)\bH^k_k\bv^*_k\bigg)^{-1}\bigg)\nonumber\\
&=\log\bigg|\mathbf{I}_{N}+\mathbf{H}^k_{k}\mathbf{v}^*_{k}(\bv^*_{k})^H({\bH}^k_{k})^H
\bigg(\sum_{\ell\ne k}
\mathbf{H}^\ell_{k}\mathbf{v}^*_{\ell}(\bv^*_{\ell})^H({\bH}^{\ell}_{k})^H+\sigma^2_{k}\mathbf{I}_N\bigg)^{-1}\bigg|\nonumber\\
&=R_k(\bv^*).\nonumber
\end{align}}\hspace{-0.2cm}
Using Definition \ref{defStationary}, we write the stationarity
condition of problem \eqref{problemWMMSESingleUnconstrained} w.r.t.
each component as{\small
\begin{align}
f_{\rm{mse}}'\left(\bv^*,\bu^*,\bw^*;
(0,\cdots,\bd_{\bv^{q_k}_k},0,\cdots,0)\right)\ge 0,\
\forall~\bd_{\bv^{q_k}_k}, ~\forall~q_k,\
\forall~k.\label{eqVStationary}
\end{align}}\hspace{-0.2cm}
Using Danskin's Theorem \cite{bertsekas99} and the fact that
$(\bw^*,\bu^*)=\arg \min  f_{\rm{mse}}(\bv^*,\bu,\bw)$, we have
{\small
\begin{align}
&f_{\rm{mse}}'(\bv^*,  \bu^*, \bw^*;
(0,\cdots,\bd_{\bv^{q_k}_k},0,\cdots,0))\nonumber\\
&= -f'_{\rm{R}}(\bv^*; (0,\cdots,\bd_{\bv^{q_k}_k},0,\cdots,0)),
\forall~\bd_{\bv^{q_k}_k}, \forall~q_k\in\mathcal{Q}_k,\
\forall~k\in\mathcal{K}.\nonumber
\end{align}}\hspace{-0.2cm}
Combining \eqref{eqVStationary} and the regularity of
$f_{\rm{R}}(\bv)$ given in Definition \ref{defRegular}, we conclude
that{\small
\begin{align}
-f'_{\rm R}(\bv^*; \bd_{\bv_k})\ge 0,~
\forall~\bd_{\bv_k}=\bigg[\bd_{\bv_k^{1_k}},\cdots,\bd_{\bv_k^{Q_k}}\bigg].
\end{align}}\hspace{-0.2cm}
According to Definition \ref{defStationary}, $\bv^*$ satisfies the
stationarity condition for problem
\eqref{problemSingleUnconstrained}. The reverse direction can be
obtained using the same argument.

The equivalence of the global optimal solutions of the two problems
can be argued as follows. Suppose $(\bv^*, \bu^*, \bw^*)$ is a
global optimal solution of $\min f_{\rm{mse}}(\bv,\bu,\bw)$ but
$\bv^*$ is not a global optimal solution of $\max f_{\rm R}(\bv)$.
Then there must exist a $\widehat{\bv}$ such that $f_{\rm
R}(\widehat{\bv})>f_{\rm R}(\bv^*)$. Using
\eqref{eqEquivalenceMMSERate}, we must have $f_{\rm{mse}}(\bv^*,
\bu^*, \bw^*)>K-f_{R}(\widehat{\bv})$. Notice that when plugging
$\widehat{\bv}$, $\bu^*(\widehat{\bv})$, $\bw^*(\widehat{\bv})$ into
$f_{\rm{mse}}(\cdot)$, we again have{\small
\begin{align}
f_{\rm{mse}}(\widehat{\bv},
\bu^*(\widehat{\bv}),\bw^*(\widehat{\bv}))=K-f_R(\widehat{\bv}).
\end{align}}
Therefore, we have{\small
\begin{align}
f_{\rm{mse}}(\bv^*,  \bu^*, \bw^*)>f_{\rm{mse}}\left(\widehat{\bv},
\bu^*(\widehat{\bv}),\bw^*(\widehat{\bv})\right),
\end{align}}
a contradiction to the global optimality of $(\bv^*, \bu^*, \bw^*)$.
This completes the second part of the claim. \hfill $\blacksquare$

{\it Proof of Proposition \ref{propEquivalenceGeneral}} (sketch) We first show that the function
$\gamma_{i_k}(\cdot)$ is well defined.
From our assumption on the utility function $u_{i_k}(\cdot)$, we see that
$u_{i_k}(-\log(e_{i_k}))$ is a strictly convex function in
${e}_{i_k}$ for all $e_{i_k}\ge 0$.
This ensures $\frac{ d -u_{i_k}(- \log(e_{i_k}))}{d e_{i_k}}$ is a strictly decreasing function.
Consequently,
its inverse function is well defined. 
Assume that $(\bv^*, \bu^*,\bw^*)$ is a stationary
solution to problem ${\rm (P2)}$. Following the steps of the proof in
Proposition \ref{propEquivalence}, we can show that $w^*_{i_k}$ is of the following form{\small
\begin{align}
w^*_{i_k}&={\frac{du_{i_k}(R_{i_k})}{dR_{i_k}}}\bigg|_{R_{i_k}=R_{i_k}(\bv^*)}\nonumber\\
&\quad\times
(1-(\bv^*_{i_k})^H(\bH^k_{i_k})^H\mathbf{C}^{-1}_{i_k}(\bv^*)
\bH^k_{i_k} \bv^*_{i_k})^{-1}.\nonumber
\end{align}}
The rest of the proof is the same as that of Proposition \ref{propEquivalence}.
We omit it due to space limit.\hfill $\blacksquare$

\section{}\label{appBisection}

We first show a monotonicity property of
$h_{i_k}(\delta_{i_k}^{q_k},\mu^{q_k})$.

\newtheorem{L2}{\bf Lemma}
\begin{L1}\label{lemmaMonotonicity}
{\it Suppose $\big\|\bc_{i_k}\big\|>\frac{\lambda_k}{2}$. Then for
fixed $\delta_{i_k}^{q_k}> 0$,
$h_{i_k}(\delta_{i_k}^{q_k},\mu^{q_k})$ is a strictly decreasing
function of $\mu^{q_k}$. For fixed $\mu^{q_k}\ge 0$,
$h_{i_k}(\delta_{i_k}^{q_k},\mu^{q_k})$ is a strictly increasing
function of $\delta_{i_k}^{q_k}$. }
\end{L1}
\begin{proof}
Define $
\bB(\delta_{i_k}^{q_k},\mu^{q_k})\triangleq\bJ_k[q,q]+\big(\frac{\lambda_k\delta_{i_k}^{q_k}}{2}+\mu^{q_k}\big)
\bI_M$. Then we have{\small
\begin{align}
&\frac{\partial h_{i_k}(\delta_{i_k}^{q_k},\mu^{q_k})}{\partial
\mu^{q_k}}=\frac{\delta_{i_k}^{q_k}}{2}\|\bB^{-1}(\delta_{i_k}^{q_k},\mu^{q_k})\bc_{i_k}\|^{-1}\nonumber\\
&\quad\quad \times
\frac{\partial\trace\left[\bB^{-1}(\delta_{i_k}^{q_k},\mu^{q_k})\bc_{i_k}\bc_{i_k}^H\bB^{-H}(\delta_{i_k}^{q_k},\mu^{q_k})\right]}
{\partial
\mu^{q_k}}\nonumber\\
&=-{\delta_{i_k}^{q_k}}\|\bB^{-1}(\delta_{i_k}^{q_k},\mu^{q_k})\bc_{i_k}\|^{-1}\nonumber\\
&\quad\quad\times
\trace\left[\bB^{-1}(\delta_{i_k}^{q_k},\mu^{q_k})\bB^{-1}(\delta_{i_k}^{q_k},\mu^{q_k})\bc_{i_k}\bc_{i_k}^H\bB^{-H}
(\delta_{i_k}^{q_k},\mu^{q_k})\right]\nonumber
\end{align}}\hspace{-0.2cm}
Notice that $\bJ_k[q,q]\succeq 0$ because it is a principal
submatrix of a positive semidefinite matrix $\bJ_k$. This fact
combined with $\delta_{i_k}^{q_k}>0$ ensures
$\bB(\delta_{i_k}^{q_k},\mu^{q_k})\succ 0$ and
$\bB^{-1}(\delta_{i_k}^{q_k},\mu^{q_k})\succ 0$. Using the fact that
$\mathbf{c}_{i_k}\ne \bf{0}$ and
$\bB^{-1}(\delta_{i_k}^{q_k},\mu^{q_k})\succ 0$, we have{\small
\begin{align}
&\trace\left[\bB^{-1}(\delta_{i_k}^{q_k},\mu^{q_k})\bB^{-1}(\delta_{i_k}^{q_k},\mu^{q_k})\bc_{i_k}\bc_{i_k}^H\bB^{-H}
(\delta_{i_k}^{q_k},\mu^{q_k})\right]\nonumber\\
&=\bc_{i_k}^H\bB^{-H}(\delta_{i_k}^{q_k},\mu^{q_k})\bB^{-1}
(\delta_{i_k}^{q_k},\mu^{q_k})\bB^{-1}(\delta_{i_k}^{q_k},\mu^{q_k})\bc_{i_k}>0.\nonumber\
\end{align}}\hspace{-0.2cm}
This condition ensures
$\frac{h_{i_k}(\delta_{i_k}^{q_k},\mu^{q_k})}{\partial \mu^{q_k}}<
0$, $\forall~\delta_{i_k}^{q_k}>0$, which in turn implies the
desired monotonicity.

The second part of the lemma can be shown similarly. 
\end{proof}

Utilizing Lemma \ref{lemmaMonotonicity}, we can show that
$\delta_{i_k}^{q_k}(\mu^{q_k})$ always exists.
\newtheorem{L3}{Lemma}
\begin{L1}\label{lemmaExistenceDelta}
{\it Suppose the condition $\big\|\bc_{i_k}\big\|>
\frac{\lambda_k}{2}$ is satisfied. Then for any fixed $\mu^{q_k}$
that satisfies $0\le \mu^{q_k}< \infty$, there always exists a
$\delta_{i_k}^{q_k}(\mu^{q_k})$ that satisfies
\eqref{eqDeltaEquality} for all $i_k\in\mathcal{A}^{q_k}$.}
\end{L1}
\begin{proof}
Pick an $i_k\in\mathcal{A}^{q_k}$. First notice that
$h_{i_k}(0,\mu^{q_k})=0,\ \forall\ 0\le \mu^{q_k}<\infty$. We then
show that
$\lim_{\delta_{i_k}^{q_k}\to\infty}h_{i_k}(\delta_{i_k}^{q_k},\mu^{q_k})>1
,\ \forall\ 0\le \mu^{q_k}<\infty$. To this end, we can write{\small
\begin{align}
&\lim_{\delta_{i_k}^{q_k}\to\infty}h_{i_k}(\delta_{i_k}^{q_k},\mu^{q_k})\nonumber\\
&=\lim_{\delta_{i_k}^{q_k}\to\infty}\frac{\delta_{i_k}^{q_k}}{\frac{\lambda_k}{2}\delta_{i_k}^{q_k}+\mu^{q_k}}
\bigg\|\bigg(\bJ_k[q,q]\frac{1}{\frac{\lambda_k}{2}\delta_{i_k}^{q_k}+\mu^{q_k}}+\bI_M\bigg)^{-1}\bc_{i_k}\bigg\|\nonumber\\
&=\frac{\|\bc_{i_k}\|}{\frac{\lambda_k}{2}}>1\nonumber
\end{align}}\hspace{-0.2cm}
where the last inequality is from the assumption.

Combining Lemma \ref{lemmaMonotonicity} and the fact that
$h_{i_k}(\delta_{i_k}^{q_k},\mu_{q_k})$ is increasing w.r.t.
$\delta_{i_k}^{q_k}$, we conclude from continuity that there must exist
$0<\delta_{i_k}^{q_k}(\mu^{q_k})<\infty$ such that
$\eqref{eqDeltaEquality}$ is satisfied.
\end{proof}
The proof of Lemma \ref{lemmaExistenceDelta} is constructive, as it
ensures that a bisection method can find
$\delta_{i_k}^{q_k}(\mu^{q_k})$.

\section{Proof of Lemma \ref{lemmaMonotonicityDelta}} \label{appMonotonicityDelta}
Fix a given $\mu^{q_k}\ge 0$, pick $i_k\in\mathcal{A}^{q_k}$. Suppose
$\delta_{i_k}^{q_k}(\mu^{q_k})$ satisfies
$h_{i_k}(\delta_{i_k}^{q_k}(\mu^{q_k}), \mu^{q_k})=1$. Fix
$\widehat{\mu}^{q_k}>\mu^{q_k}$. From Lemma \ref{lemmaMonotonicity}
we have 
$h_{i_k}(\delta_{i_k}^{q_k}(\mu^{q_k}), \widehat{\mu}^{q_k})< 1$.
To ensure
$h_{i_k}(\delta_{i_k}^{q_k}(\widehat{\mu}^{q_k}), \widehat{\mu}^{q_k})=1$,
 we must have $\delta_{i_k}^{q_k}(\widehat{\mu}^{q_k})>
\delta_{i_k}^{q_k}({\mu}^{q_k})$, which gives the first part of the claim. 

We prove the second part of the claim by contradiction. First
consider the trivial case where no user is active, i.e.,
$\mathcal{A}^{q_k}=\emptyset$. Clearly
$\sum_{i_k\in\mathcal{I}_k}\|\bv^{q_k}_{i_k}(\mu^{q_k})\|^2=0$ and the claim is proven.

Now suppose $\mathcal{A}^{q_k}$ is nonempty, that is, $A^{q_k}>0$.
Suppose that there exists an $i_k\in\mathcal{A}_k$ such that for all
$\mu^{q_k}\ge 0$,
$\delta_{i_k}^{q_k}(\mu^{q_k})<\sqrt{\frac{{A}^{q_k}}{P_{q_k}}}$.
This assumption combined with Lemma \ref{lemmaMonotonicity} implies{\small
\begin{align}
h_{i_k}\bigg(\sqrt{\frac{{A}^{q_k}}{P_{q_k}}},\mu^{q_k}\bigg)>h_{i_k}(\delta_{i_k}^{q_k}(\mu^{q_k}),\mu^{q_k})=1,
\ \forall \ \mu^{q_k}\ge 0.\label{eqContradictionH}
\end{align}}\hspace{-0.2cm}
However, we have that{\small
\begin{align}
&\lim_{\mu^{q_k}\to\infty}h_{i_k}\bigg(\sqrt{\frac{{A}^{q_k}}{P_{q_k}}},\mu^{q_k}\bigg)\nonumber\\
&=\lim_{\mu^{q_k}\to\infty}\frac{\sqrt{P_{q_k}}}{\frac{\lambda_k}{2}\sqrt{{A}^{q_k}}+\mu^{q_k}\sqrt{P_{q_k}}}
\bigg\|\bigg(\bJ_k[q,q]\frac{1}{\frac{\lambda_k
\sqrt{{A}^{q_k}}}{2\sqrt{P_{q_k}}}+\mu^{q_k}}+\bI_M\bigg)^{-1}\bc_{i_k}\bigg\|\nonumber\\
&=0\nonumber
\end{align}}\hspace{-0.2cm}
which contradicts \eqref{eqContradictionH}. This suggests that for
each $i_k\in\mathcal{A}^{q_k}$, there exists a
$\overline{\mu}^{q_k}_{i_k}$ such that for all $\mu^{q_k}\ge
\overline{\mu}^{q_k}_{i_k}$,
$\delta_{i_k}^{q_k}(\mu^{q_k})\ge\sqrt{\frac{{A}^{q_k}}{P_{q_k}}}$
(i.e.,
$\frac{1}{(\delta_{i_k}^{q_k}(\mu^{q_k}))^2}\le\frac{P_{q_k}}{{A}^{q_k}}$).
Taking
$\overline{\mu}^{q_k}=\max_{i_k\in\mathcal{A}^{q_k}}\overline{\mu}^{q_k}_{i_k}$,
we have that for all $\mu^{q_k}\ge \overline{\mu}^{q_k}$,
$\sum_{i_k\in\mathcal{A}^{q_k}}\frac{1}{(\delta_{i_k}^{q_k}(\mu^{q_k}))^2}\le{P_{q_k}}$.
This condition implies that {\small$
\sum_{i_k\in\mathcal{I}_k}\|\bv^{q_k}_{i_k}(\mu^{q_k})\|^2\le
P_{q_k}\nonumber.$} As a result, the second part of the claim is proved.

\vspace{-0.2cm}
\section{Proof of Theorem \ref{theoremConvergenceVSingle}}\label{appConvergence}
\begin{proof}
Due to the equivalence relationship, it is sufficient to show that
the S-WMMSE algorithm converges to a stationary solution of the problem ${\rm (P2)}$.

We first show that the BCD procedure in Table \ref{tableVUpdate} for updating $\bv$ converges to the
global optimal solution of problem {\rm (P3)}. Recall that
this problem can be decomposed into $K$ independent convex subproblems of the form
{\rm (P4)}, then it is sufficient to show that each of these problems
are solved globally. Similarly as
\eqref{problemSingleUnconstrained}-\eqref{problemWMMSESingleUnconstrained},
problem {\rm (P4)} can be expressed in its unconstrained form
{\small
\begin{align}
\min_{\{\bv_{i_k}\}_{i_k\in\mathcal{I}_k}}&\sum_{i_k\in\mathcal{I}_k}\bigg(\bv^H_{i_k}
\bJ_k\bv_{i_k}-\bv^H_{i_k}\bd_{i_k}-\bd^H_{i_k}\bv_{i_k}\nonumber\\
&\quad
+\lambda_k\sum_{q_k\in\mathcal{Q}_k}\|\bv^{q_k}_{i_k}\|-\sum_{q_k\in\mathcal{Q}_k}I(\bv^{q_k}_{i_k})\bigg)\nonumber.
\end{align}}\hspace{-0.2cm}
The procedure in Table \ref{tableVUpdate} is a BCD method for
solving the above unconstrained problem, where each block is defined
as $\bv^{q_k}\triangleq\{\bv^{q_k}_{i_k}\}_{i_k\in\mathcal{I}_k}$.
Observe that {\it i)} the nonsmooth part of the objective is {\it
separable} across the blocks; {\it ii)} the smooth part of the
objective is differentiable; {\it iii)} each block variable
$\bv^{q_k}$ can be solve uniquely when fixing other variables
$\{\bv^{p_k}\}_{p_k\ne q_k}$. According to \cite[Theorem
4.1-(c)]{tseng01}, these facts are sufficient to guarantee the
convergence of this BCD procedure to a global optimal solution of
the convex nonsmooth problem {\rm (P4)}.

To prove the convergence of the S-WMMSE algorithm to a stationary solution of problem ${\rm (P2)}$,
we can again write problem ${\rm (P2)}$ into its unconstrained form, 
and see that the nonsmooth part of the objective is {separable} across the blocks
of variables $\bv, \bu, \bw$. Furthermore,
when we fix any two block variables and solve for the third,
a {unique} optimal solution can be obtained.
Applying \cite[Theorem 4.1]{tseng01}, we conclude that the S-WMMSE algorithm converges to
a stationary solution of the problem {\rm (P2)}.
\end{proof}


\bibliographystyle{IEEEbib}
{\footnotesize
\bibliography{ref}
}

\end{document}